\documentclass[%
reprint,
superscriptaddress,
 amsmath,amssymb,
 aps,
 prl,
floatfix,
]{revtex4-2}

\usepackage{url}

\usepackage{graphicx}
\usepackage{dcolumn}
\usepackage{bm}
\usepackage{color}

\usepackage[colorlinks=true, urlcolor=blue, linkcolor=blue, citecolor=blue, pdftex]{hyperref}

\usepackage{braket}
\usepackage{color}
\usepackage{epstopdf}

\newcommand{\BCTO}{$\mathrm{Ba}_2\mathrm{CoTeO}_6$}
\newcommand{\BCSO}{$\mathrm{Ba}_3\mathrm{CoSb}_2\mathrm{O}_9$}

\begin{document}

\preprint{APS/123-QED}

\title{Magnons and Spinons in $\mathbf{Ba_2CoTeO_6} $: A Composite System of Isolated Spin-$\bm{1/2}$ Triangular Heisenberg-like and Frustrated Honeycomb Ising-like Antiferromagnets}

\author{Yuki Kojima}
\email{kojima.y.ai@m.titech.ac.jp}
\author{Nobuyuki Kurita}
\author{Hidekazu Tanaka}
\email{tanaka@lee.phys.titech.ac.jp}
\affiliation{
Department of Physics, Tokyo Institute of Technology, Meguro, Tokyo 152-8551, Japan}
\author{Kenji Nakajima}
\affiliation{
Materials and Life Science Division, J-PARC Center, Tokai, Ibaraki 319-1195, Japan}
\affiliation{
Materials Sciences Research Center, Japan Atomic Energy Agency, Sayo, Hyogo 679-5148, Japan}

\date{\today}

\begin{abstract}
We report the neutron scattering results on magnetic orderings and excitations in {\BCTO} composed of two almost isolated subsystems A and B, which are described as an $S\,{=}\,1/2$ triangular Heisenberg-like antiferromagnet and a frustrated honeycomb Ising-like antiferromagnet, respectively. Stripy ordering of subsystem B was confirmed below $T_{\rm N1}\,{=}\,12.0$\,K, whereas sharp streaks were observed along $(1/3, 1/3, L)$ and $(2/3, 2/3, L)$ at 0.3\,K (${\ll}\,T_{\rm N2}\,{=}\,3.0$\,K). This indicates the two-dimensional nature of ordering in subsystem A. It was found that the excitation spectra of both subsystems are well separated and independent of each other. The excitation spectrum of subsystem A is composed of two single-magnon branches with roton-like minima at the M point and a clearly structured intense continuum, as similarly observed in {\BCSO}, which is strongly indicative of spinon excitations. Dispersion curves for subsystem B can be described by linear spin wave theory within the third-neighbor exchange interaction. 
\end{abstract}


\maketitle

 
The ground state of a two-dimensional (2D) spin-1/2 triangular-lattice Heisenberg antiferromagnet (TLHAF) with the nearest-neighbor exchange interaction at zero magnetic field has been considered to be an ordered state with the 120$^{\circ}$ structure~\cite{Huse1988,Jolicoeur1989,Bernu1994,Singh1992,White2007,Gotze2016} rather than a spin liquid state as the resonating-valence-bond (RVB) state~\cite{Anderson1973}. However, the problem whether the RVB state vanishes completely or survives in the ground state of the $S\,{=}\,1/2$ TLHAF has not been resolved, although it seems that the ordered state and the RVB state are mutually exclusive. A study of static magnetic properties appears to be insufficient to answer this problem; thus, a study of magnetic excitations is necessary. Note that the $S\,{=}\,1/2$ TLHAF exhibits a wide magnetization plateau at one-third of the saturation magnetization stabilized by quantum fluctuation~\cite{Chubukov1991,Nikuni1993,Honecker1999,Alicea2009,Farnell2009,Sakai2011,Hotta2013,Yamamoto2014,Starykh2015,Sellmann2015,Coletta2016}.

In contrast to the ground state properties, the magnetic excitations in the $S\,{=}\,1/2$ TLHAF are less well understood. The theoretical consensus appears to be limited for single-magnon excitations as follows: the excitation energy is significantly renormalized downward in a large area of the Brillouin zone, although the dispersion relation of low-energy excitations near the K point is described by linear spin wave theory (LSWT)~\cite{Starykh2006,Zheng2006,Chernyshev2009,Mezio2011,Mourigal2013,Ghioldi2015}. In addition, the dispersion curve shows a roton-like minimum at the M point, the origin of which has been interpreted to be spinon--antispinon pairs with each having spin-1/2~\cite{Zheng2006}. However, the theoretical consensus for the excitation continuum that reflects the characteristics of magnetic quasiparticles is also limited.

The predictions for the single-magnon excitations were partly confirmed by inelastic neutron scattering (INS) experiments on {\BCSO}~\cite{Ma2016,Ito2017,Kamiya2018,Macdougal2020}, which approximates an $S\,{=}\,1/2$ TLHAF~\cite{Shirata2012,Zhou2012,Susuki2013,Quirion2015,Koutroulakis2015,Yamamoto2015,Li2019,Liu2019a,Fortune2021}. However, the observed intense structured excitation continuum extending to high energies was markedly different from theory~\cite{Ghioldi2015,Kamiya2018}. The features of an excitation spectrum, which cannot be described by LSWT, appeared to be strongly indicative of fractionalized excitations. The results of INS experiments on {\BCSO} motivated the theoretical studies of magnetic excitations based on spinon excitations~\cite{Ghioldi2018,Zhang2020,Ferrari2019,Verresen2019}. In particular, approaches based on the RVB-like state with spinon excitations successfully reproduced the dispersion minima at both M and Y points~\cite{Ferrari2019,Zhang2020}. These theoretical results demonstrated that the inherent quantum fluctuations still retain the characteristics of the RVB state even in the ordered ground state.

{\BCSO} shows weak interlayer exchange interactions leading to 3D ordering. Thus, there is a possibility that its excitation spectrum is affected by the interlayer exchange interactions. For the experimental establishment of a universal excitation spectrum for the $S\,{=}\,1/2$ TLHAF, we need a model system with better two-dimensionality than {\BCSO}.
In this letter, we show that an almost isolated layer of the $S\,{=}\,1/2$ TLHAF is realized in {\BCTO}, and that its excitation spectrum is universal in the $S\,{=}\,1/2$ TLHAF.

\begin{figure}[t]
	\centering
	\includegraphics[width=8 cm, clip]{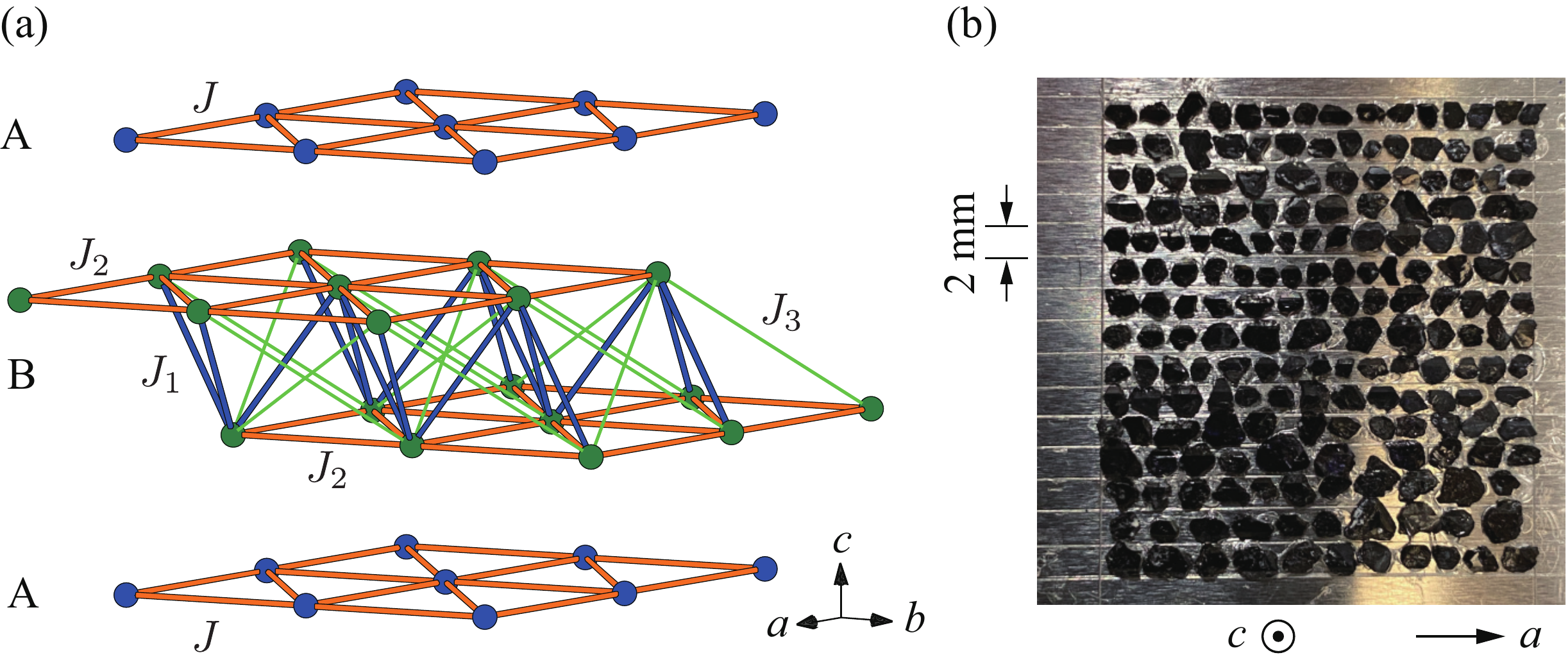}
	\caption{(Color online) (a) Magnetic subsystems A and B in {\BCTO}. Thick blue and orange lines and thin green lines in subsystem B depict exchange interactions $J_1$, $J_2$ and $J_3$, respectively. (b) Photograph of {\BCTO} single crystals co-aligned on an aluminum plate for the present INS experiment. The wide plane is the crystallographic $c$ plane. The $a$ axis is parallel to the horizontal direction.}
	\label{fig:Exchange_Crystal}
\end{figure}

{\BCTO} crystallizes in a trigonal structure with the space group $P{\bar 3}m1$~\cite{Ivanov2010}. {\BCTO} is a unique antiferromagnet composed of two highly frustrated subsystems A and B, as shown in Fig.~\ref{fig:Exchange_Crystal}\,(a)~\cite{Chanlert2016}. Subsystem A is expressed as an $S\,{=}\,1/2$ TLHAF with small XY-like exchange anisotropy. Subsystem B is composed of two uniform triangular lattices of $\mathrm{Co}^{2+}$ ions with effective spin-1/2. Because these two triangular lattices are strongly coupled by exchange interactions $J_1$ and $J_3$, and $\mathrm{Co}^{2+}$ is affected by a strong trigonal crystal field due to the distortion of a $\mathrm{CoO}_6$ octahedron, subsystem B is described as a frustrated honeycomb-lattice Ising-like antiferromagnet when projected onto the $ab$ plane. {\BCTO} undergoes successive magnetic phase transitions at $T_{\rm N1}\,{=}\,12.0$ and $T_{\rm N2}\,{=}\,3.0$\,K, which correspond to the orderings of subsystems B and A, respectively~\cite{Chanlert2016}. Previous magnetic and thermodynamic measurements and electron spin resonance (ESR) experiments indicated that these two subsystems are almost decoupled~\cite{Chanlert2016,Chanlert2017}. Because subsystems A and B have different energy scales~\cite{Chanlert2016,Chanlert2017}, it is expected that the excitation spectra of these two subsystems will be observed separately, as shown below. 

Single crystals of {\BCTO} with a typical size of $2\,{\times}\,2{\times}\,0.3$ mm$^3$ were grown by the same procedures as described in Ref.~\cite{Chanlert2016}. Single-crystal INS experiments were performed using the cold-neutron disk chopper spectrometer AMATERAS installed in the Materials and Life Science Experimental Facility (MLF) at the Japan Proton Accelerator Research Complex (J-PARC)~\cite{Nakajima2011}. Approximately 400 pieces of single crystals weighing about 1.3\,g were co-aligned on two aluminum plates along the $(1,\,1,\,0)$ and $(0,\,0,\,1)$ directions in the horizontal plane as shown in Fig.~\ref{fig:Exchange_Crystal}\,(b), using hydrogen-free CYTOP$^\circledR$ as the glue. The mosaic spread was about 3$^{\circ}$. 
The sample was cooled to 0.3\,K using a $^3$He refrigerator. Scattering data were collected by rotating the sample around the vertical $(-1,1,0)$ direction by the multi-$E_{\rm i}$ technique with a set of incident neutron energies $E_{\rm i}\,{=}\,3.6, 5.6, 9.7$ and 20.9\,meV. Elastic energy resolutions are given by $\Delta E_{\rm i}\,{=}\,0.14, 0.3, 0.6$ and 1.8\,meV, respectively. All the data were analyzed using the software suite UTSUSEMI~\cite{Inamura2013}.

\begin{figure}[t]
	\centering
	\includegraphics[width=8 cm, clip]{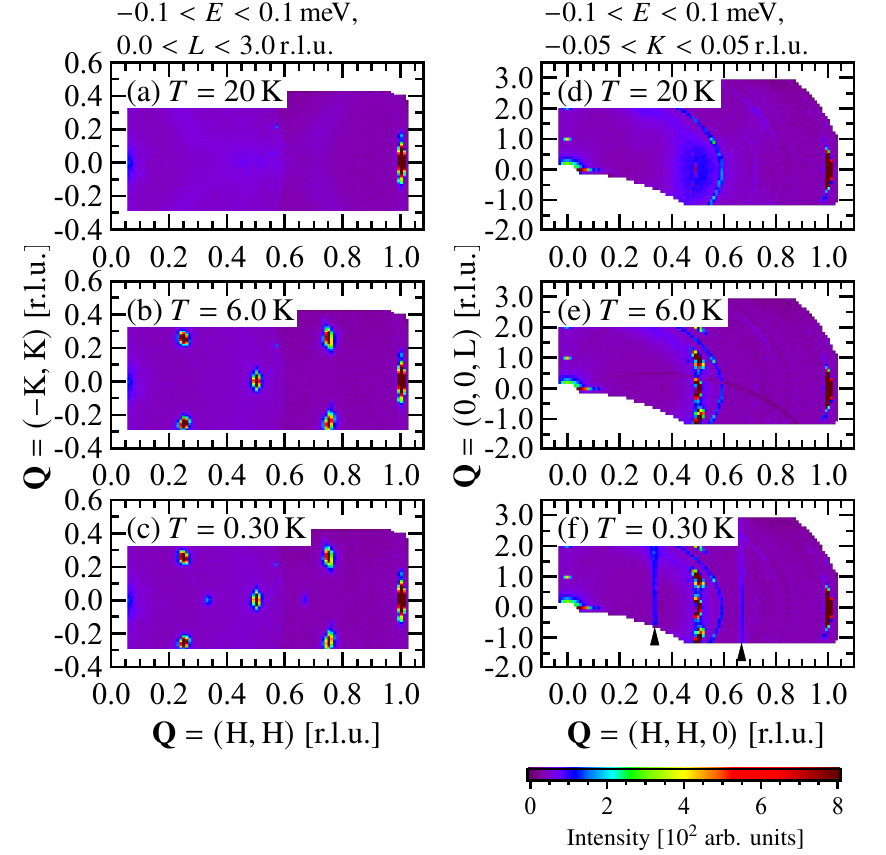}
	\caption{(Color online) (a)\,{--}\,(c) ENS intensity maps in the 2D reciprocal lattice measured with $E_{\rm i}\,{=}\,3.6$\,meV at $T\,{=}\,20$\,K (${>}\,T_{\rm N1}$), 6.0\,K ($T_{\rm N1}\,{>}\,T\,{>}\,T_{\rm N2}$) and 0.30\,K (${\ll}\,T_{\rm N2}$), respectively. Integrated energy and momentum ranges are shown in the figures. (d)\,{--}\,(f) ENS intensity maps in the $(H,\,H,\,L)$ plane, where the scattering intensity was integrated for $-0.05\,{\leq}\,K\,{\leq}\,0.05$ along $\bm Q\,{=}\,(-K,\,K,\,0)$. Strong intensities around $\bm Q\,{=}\,(0,\,0)$ due to the direct beam are masked in (a)\,{--}\,(f). Two triangles in (f) indicate diffuse streak patterns along $\bm Q\,{=}\,(1/3,\,1/3,\,L)$ and $(2/3,\,2/3,\,L)$. Curved streaks in (d)\,{--}\,(f) arise from a nonmagnetic origin.}
	\label{fig:Elastic_vsT}
\end{figure}

\begin{figure*}[t]
	\centering
	\includegraphics[width=16 cm, clip]{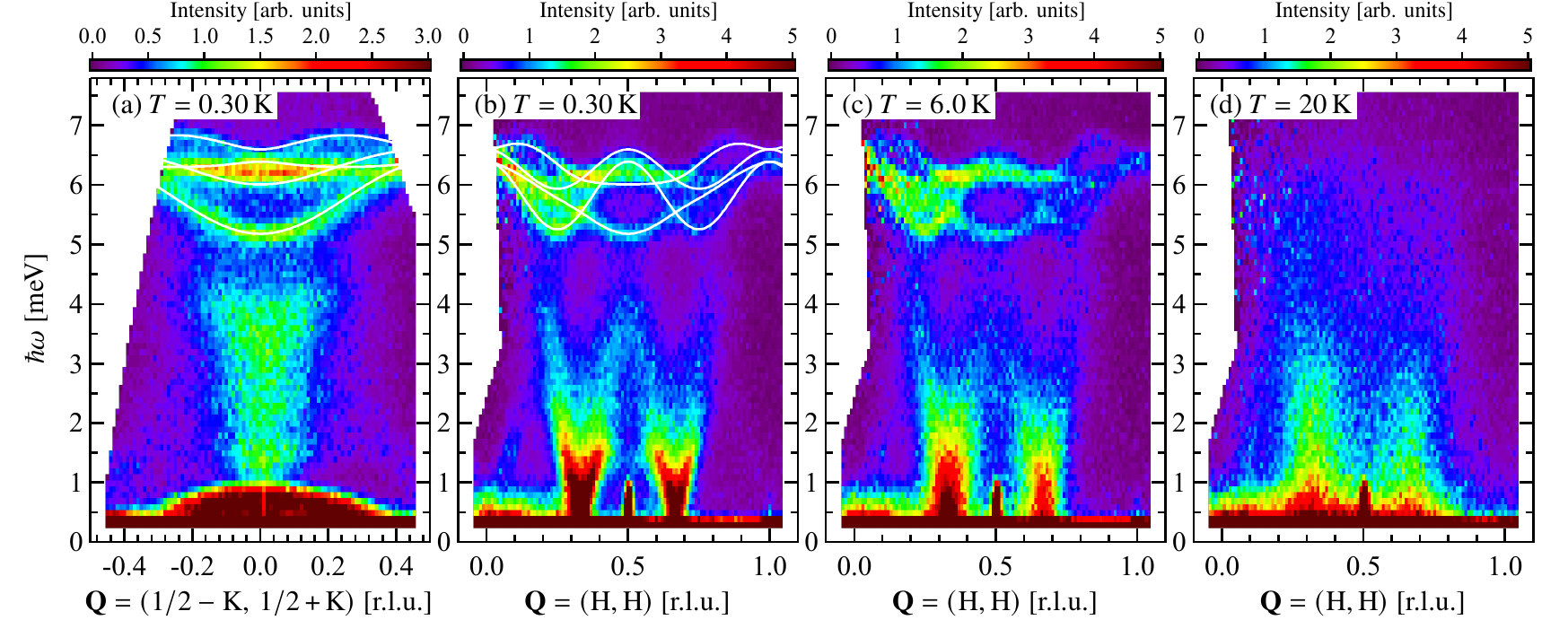}
	\caption{(Color online) (a) and (b) Excitation spectra of {\BCTO} measured with $E_{\rm i}\,{=}\,9.7$\,meV at $T\,{=}\,0.3$\,K along two high-symmetry directions ${\bm Q}\,{=}\,(1/2{-}K,\,1/2{+}K)$ and $(H,\,H)$, respectively. (c) and (d) Thermal evolution of excitation spectrum along ${\bm Q}\,{=}\,(H,\,H)$ measured at $T\,{=}\,6.0$ and 20\,K, respectively. The scattering intensities were integrated in a range of $-3\,{\leq}\,L\,{\leq}\,3$ to map the scattering intensity in the 2D reciprocal lattice assuming good two-dimensionality. The solid lines in (a) and (b) are dispersion curves of Ising-like subsystem B calculated on the basis of LSWT using the $J_1{-}J_2{-}J_3$ XXZ model of Eq.~(\ref{model_B}) with the exchange parameters shown in the text.}
	\label{fig:ExcitationSpectram_vsT}
\end{figure*}

Figures~\ref{fig:Elastic_vsT}\,(a)\,{--}\,(c) show the intensity maps of elastic neutron scattering (ENS) in the 2D reciprocal lattice plane measured with $E_{\rm i}\,{=}\,3.6$\,meV at $T\,{=}\,20$\,K (${>}\,T_{\rm N1}\,{=}\,12.0$\,K), 6.0\,K ($T_{\rm N1}\,{>}\,T\,{>}\,T_{\rm N2}\,{=}\,3.0$\,K) and 0.30\,K (${\ll}\,T_{\rm N2}$), respectively, where the scattering intensity was integrated in a momentum range of $0\,{\leq}\,L\,{\leq}\,3$. Figures~\ref{fig:Elastic_vsT}\,(d)\,{--}\,(f) show the ENS intensity maps in the $(H,\,H,\,L)$ plane measured at $T\,{=}\,20$, 6.0 and 0.3\,K, respectively. The strong Bragg peak commonly observed at ${\bm Q}\,{=}\,(1,\,1,\,0)$ is a nuclear peak. 
As the temperature decreases below $T_{\rm N1}\,{=}\,12.0$\,K, magnetic Bragg peaks emerge at ${\bm Q}\,{=}\,(1/2,\,0,\,0)$ and its equivalent positions, which is consistent with the results of a previous study by Ivanov {\it et al.}~\cite{Ivanov2010}. These magnetic Bragg peaks are attributed to a stripy order of Ising-like subsystem B characterized by the propagation vector ${\bm k}\,{=}\,(1/2, 0, 0)$. 

With a further decrease in temperature to $T\,{=}\,0.3$\,K, which is much lower than $T_{\rm N2}\,{=}\,3.0$\,K, sharp streak patterns are observed along $\bm Q\,{=}\,(1/3,\,1/3,\,L)$ and $(2/3,\,2/3,\,L)$, as shown in Fig.~\ref{fig:Elastic_vsT}\,(f). After integrating the scattering intensity over $L$, these streaks are observed as sharp spots in the 2D reciprocal lattice, as shown in Fig.~\ref{fig:Elastic_vsT}\,(c). This result suggests that below $T_{\rm N2}$, the 120$^{\circ}$ structure is constructed in a triangular layer of subsystem A, whereas the interlayer ordering is not long-range but short-range. This is an unexpected result, because specific heat displays a sharp peak at $T_{\rm N2}$, indicative of the 3D ordering. 

\begin{figure}[t]
	\centering
	\includegraphics[width=8.5 cm, clip]{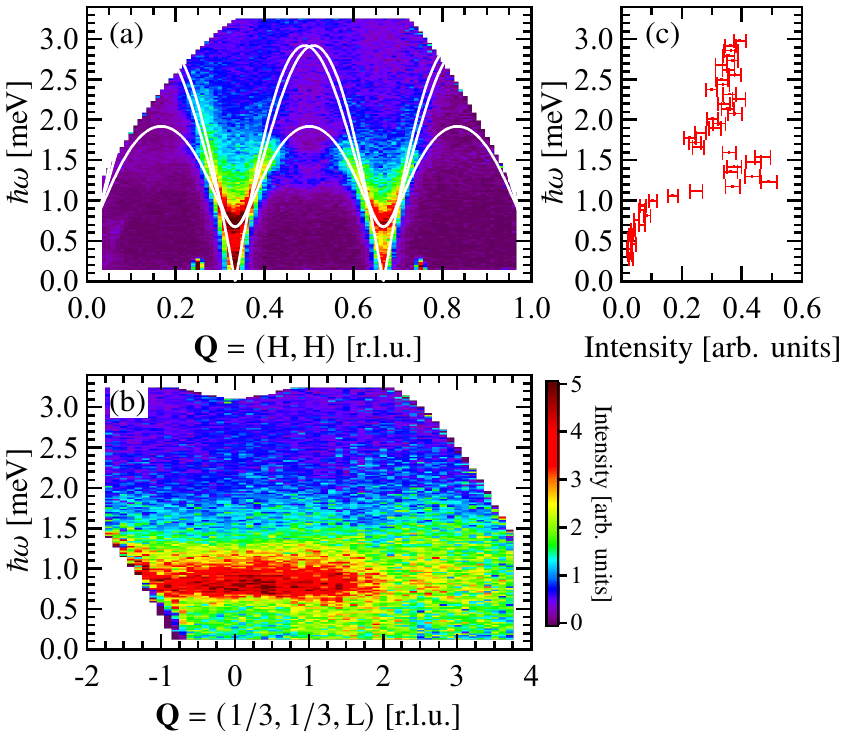}
	\caption{(Color online) Energy-momentum map of low-energy excitations of {\BCTO} measured with $E_{\mathrm{i}}\,{=}\,3.6$\,meV at $T\,{=}\,0.3$\,K along (a) ${\bm Q}\,{=}\,(H,\,H)$, where the scattering intensities were integrated over $L$, and (b) ${\bm Q}\,{=}\,(1/3,\,1/3,\,L)$. The solid lines in (a) are LSWT dispersion curves calculated with $J\,{=}\,1.85$\,meV and ${\Delta}\,{=}\,0.040$.
(c) Scattering intensity at ${\bm Q}\,{=}\,(1/2,\,1/2)$ as a function of energy. }
	\label{fig:Excite_TL}
\end{figure}

Figures~\ref{fig:ExcitationSpectram_vsT}\,(a) and (b) show the excitation spectra of {\BCTO} in two high-symmetry directions ${\bm Q}\,{=}\,(1/2{-}K,\,1/2{+}K)$ and $(H,\,H)$, respectively, measured with $E_{\rm i}\,{=}\,9.7$\,meV at $T\,{=}\,0.3$\,K. The scattering intensities were integrated in a range of $-3\,{\leq}\,L\,{\leq}\,3$ to map the scattering intensity in the 2D reciprocal lattice. We confirmed good two-dimensionality from the observation that the excitation spectrum is dispersionless along the $c^*$ direction, as shown in Fig.~\ref{fig:Excite_TL}\,(b). The structured excitation spectrum at $T\,{=}\,0.3$\,K becomes very indistinct at $T\,{=}\,20$\,K, as shown in Fig.~\ref{fig:ExcitationSpectram_vsT}\,(d). The scattering intensity at around 6\,meV decreases rapidly at $T\,{=}\,20$\,K. This result verifies that the origins of the excitation spectra shown in Figs.~\ref{fig:ExcitationSpectram_vsT} are magnetic.

The excitation spectrum at $T\,{=}\,0.3$\,K can be divided into two parts, low-energy excitation (${\hbar}{\omega}\,{<}\,5$\,meV) and high-energy excitation ($5\,{<}\,{\hbar}{\omega}\,{<}\,7$\,meV). In previous studies~\cite{Chanlert2016,Chanlert2017}, the predominant exchange interactions were estimated as $J\,{\simeq}\,1.8$\,meV for subsystem A and $J_1\,{\simeq}\,4.8$ and $J_1\,{\simeq}\,2.4$\,meV for subsystem B, where the exchange constants were estimated on the basis of the $J_1{-}J_2$ honeycomb-lattice Ising antiferromagnet without XY components. Thus, we can deduce that the low- and high-energy spectra originate in subsystems A and B, respectively. This can also be confirmed from the temperature evolution of the excitation spectrum shown in Figs.~\ref{fig:ExcitationSpectram_vsT}\,(c) and (d). At $T\,{=}\,6.0$\,K, which is twofold $T_{\rm N2}$, the low-energy spectrum becomes indistinct, whereas the high-energy spectrum remains unchanged. These results show that the low- and high-energy spectra have different sources, which exhibit magnetic orderings separately at $T_{\rm N2}$ and $T_{\rm N1}$, respectively. In a previous magnetization study of {\BCTO}, successive transitions with a sharp stepwise magnetization curve characteristic of the ordered state in an Ising-like antiferromagnet were observed at $T\,{=}\,4.2$\,K ($T_{\rm N2}\,{<}\,T\,{<}\,T_{\rm N1}$) for magnetic fields parallel to the $c$ axis~\cite{Chanlert2016}. The result of the magnetization measurements confirms that $T_{\rm N1}$ corresponds to the ordering temperature of Ising-like subsystem B. From the energy scales and temperature evolution of the excitation spectrum, the sources of the low- and high-energy spectra can be subsystems A and B, respectively. Note that the excitation spectra of subsystem B show no change through $T_{\rm N2}$. Given that subsystems A and B are exchange-coupled, the change in the dispersion curves of subsystem B will be observed above and below $T_{\rm N2}$. Thus, we can confirm that subsystems A and B are almost decoupled.

Figure~\ref{fig:Excite_TL}\,(a) shows the spectrum of low-energy excitations from subsystem A along ${\bm Q}\,{=}\,(H,\,H)$ measured at $T\,{=}\,0.30$\,K with $E_{\mathrm{i}}\,{=}\,3.6$\,meV, where the scattering intensities were integrated in a range of $0\,{\leq}\,L\,{\leq}\,3$ assuming good two-dimensionality. Figure~\ref{fig:Excite_TL}\,(b) shows the spectrum of low-energy excitations from subsystem A along ${\bm Q}\,{=}\,(1/3,\,1/3,\,L)$ measured at $T\,{=}\,0.30$\,K with $E_{\mathrm{i}}\,{=}\,3.6$\,meV. It is notable that the excitations are completely dispersionless along ${\bm Q}\,{=}\,(1/3,\,1/3,\,L)$ compared with those from {\BCSO}, in which low-energy single-magnon excitations are dispersive along ${\bm Q}\,{=}\,(1/3,\,1/3,\,L)$ owing to the small interlayer interactions~\cite{Ito2017}. A strong dispersionless excitation with a gap of ${\hbar}{\omega}\,{=}\,0.68$\,meV is caused by the small XY-like exchange anisotropy. This result, together with the elastic scattering result shown in Fig.~\ref{fig:Elastic_vsT}\,(f), demonstrates that each triangular layer of subsystem A is almost isolated. Thus, the $S\,{=}\,1/2$ TLHAF realized in {\BCTO} has better two-dimensionality than that in {\BCSO}. 

Here we compare single-magnon excitations in subsystem A with LSWT. We consider the 2D Heisenberg model with a weak XY anisotropy expressed as
\begin{equation}
\mathcal{H}_{\rm A}=J\sum_{\braket{i,j}}\left(\bm S_i\cdot\bm S_j-\Delta S^z_iS^z_j\right),
\label{model_A}
\end{equation}
where $J$ and $\Delta$ are the coupling constant and the anisotropy parameter in subsystem A, respectively. The LSWT dispersion curves with $J\,{=}\,1.85$\,meV and ${\Delta}\,{=}\,0.040$ describe well the experimental dispersion curves below 1\,meV in the vicinity of the K point at ${\bm Q}\,{=}\,(1/3, 1/3)$ and $(2/3, 2/3)$, as shown Fig.~\ref{fig:Excite_TL}\,(a). These exchange parameters also coincide with those estimated from the saturation field of subsystem A and the zero-field gap obtained by ESR measurements~\cite{Chanlert2016,Chanlert2017}. 
The further the wave vector moves away from the K point, the more rapidly the excitation energy deviates downward from the LSWT dispersion, as predicted by theory~\cite{Starykh2006,Zheng2006,Chernyshev2009,Mezio2011,Mourigal2013,Ghioldi2015,Ghioldi2018,Zhang2020,Ferrari2019} and as observed in {\BCSO}~\cite{Ito2017,Macdougal2020}. It is notable that two single-magnon dispersion curves display roton-like minima at ${\bm Q}\,{=}\,(1/2, 1/2)$. Figure~\ref{fig:Excite_TL}\,(c) shows scattering intensity at ${\bm Q}\,{=}\,(1/2,\,1/2)$ as a function of energy. The dispersion minima with energies of ${\hbar\omega}\,{=}\,1.25$ and $1.55$\,meV correspond to the minima at the M and Y points in the single branch, respectively.

Overall, the features of the excitation spectrum of subsystem A are very similar to those observed in {\BCSO}~\cite{Ito2017,Macdougal2020}. Downward renormalized single-magnon excitation energies, two roton-like minima of the dispersion curves at ${\bm Q}\,{=}\,(1/2, 1/2)$ and intense structured excitation continuum extending to at least 5\,meV, which is approximately three times larger than that in the exchange interaction $J$ shown in Figs.~\ref{fig:ExcitationSpectram_vsT}\,(a) and (b), are the common observations. From these observations, we can deduce that the low-energy part of the excitation spectrum of {\BCTO} and the excitation spectrum of {\BCSO} are universal in the $S\,{=}\,1/2$ TLHAFs. The difference between {\BCTO} and {\BCSO} is that the intensities of single-magnon excitations around ${\bm Q}\,{=}\,(1/2, 1/2)$ in {\BCTO} are considerably weaker than those around the K point. We speculate that this spectral difference between these two systems arises from the difference in their interlayer interactions. 

There have been efforts to theoretically explain the roton-like minima ${\bm Q}\,{=}\,(1/2, 1/2)$~\cite{Starykh2006,Zheng2006,Chernyshev2009,Mezio2011,Mourigal2013,Ghioldi2015,Ghioldi2018,Zhang2020,Ferrari2019,Verresen2019}. The series expansion method can be used to derive the minimum at the M point but not at the Y point~\cite{Zheng2006}. On the other hand, approaches based on the RVB-like state with spinon excitations successfully reproduced the dispersion minima at both the M and Y points~\cite{Ferrari2019,Zhang2020}. The RVB-based approaches also roughly describe the features of the excitation continuum observed in {\BCTO} and {\BCSO}; these cannot be described by the spin wave theory~\cite{Kamiya2018}. For these reasons, it is deduced that the inherent quantum fluctuations still have the characteristics of the RVB state even in the ordered ground state. 

Next, we analyze the dispersion curves of the high-energy single-magnon excitations in an energy range of $5\,{<}\,{\hbar}{\omega}\,{<}\,7$\,meV, which originate in subsystem B. 
Because the observed spectrum is dispersive, which indicates the finite XY component, we consider the following XXZ model on a honeycomb lattice with up to the third neighbors $J_1$, $J_2$ and $J_3$ exchange interactions, as shown in Fig.~\ref{fig:Exchange_Crystal}\,(a): 
\begin{equation}
\mathcal{H}_{\rm B}=\sum_{\nu=1}^3\sum_{\braket{i,j}_\nu}J_\nu\left[S^z_iS^z_j +{\delta}(S^x_iS^x_j+S^y_iS^y_j)\right],
\label{model_B}
\end{equation}
where $\braket{ij}_\nu$ is the $\nu$-th nearest-neighbor coupling pairs. The $z$ axis is taken to be along the crystallographic $c$ axis. The anisotropy parameter ${\delta}$ is assumed to be the same for all interactions because the exchange anisotropy is generally determined by the distortion of a single $\mathrm{CoO}_6$ octahedron. 

In the previous magnetization measurements in {\BCTO}~\cite{Chanlert2016}, magnetization plateaus at one-third and one-half of the saturation magnetization were observed for magnetic fields parallel to the $c$ axis. Two critical fields $H_{\rm c1}$ and $H_{\rm c2}$ for transitions to the 1/3- and 1/2-magnetization plateau states and the saturation field $H_{\rm s}$ were obtained to be $H_{\rm c1}\,{=}\,12.3$, $H_{\rm c2}\,{=}\,14.8$ and $H_{\rm s}\,{=}\,39.0$\,T. These transition fields were analyzed within the context of a $J_1{-}J_2$ Ising model. Here, we consider the third-neighbor ferromagnetic exchange interaction $J_3$, because the dispersion curves of subsystem B cannot be reproduced within the $J_1{-}J_2$ model. Within the Ising model, the critical fields and saturation field are expressed as $H_{\rm c1}\,{=}\,J_1\,{-}\,2J_3$, $H_{\rm c2}\,{=}\,{-}(1/2)J_1\,{+}\,3J_2\,{+}\,4J_3$ and $H_{\rm s}\,{=}\,(3/2)J_1\,{+}\,3J_2$, where $g_{\rm B}^{\parallel}{\mu}_{\rm B}$ is set to be unity for simplicity. The ferromagnetic $J_3$ interaction stabilizes the stripy order at zero magnetic field, which enhances the energy gap, and destabilizes the 1/3-magnetization plateau state to reduce its field range. We calculate the dispersion curves of subsystem B on the basis of LSWT under a constraint of $H_{\rm c1}/H_{\rm c2}\,{=}\,12.3/14.8$, which leads to $J_3\,{=}\,0.266J_1\,{-}\,0.468J_2$. We assume three types of magnetic domain with the stripy order. Solid lines in Figs.~\ref{fig:ExcitationSpectram_vsT}\,(a) and (b) are dispersion curves calculated with $J_1\,{=}\,2.94$, $J_2\,{=}\,2.37$, $J_3\,{=}\,{-}0.328$\,meV and ${\delta}\,{=}\,0.15$ and are in good agreement with experimental results. Although there are generally six excitation branches owing to three magnetic domains, only four branches are observed along two high-symmetry directions ${\bm Q}\,{=}\,(1/2{-}K,\,1/2{+}K)$ and $(H,\,H)$ because two domains are equivalent. 

To conclude, we have presented excitation spectra of {\BCTO} composed of two subsystems A and B, which are described as an $S\,{=}\,1/2$ TLHAF with small XY-like exchange anisotropy and a frustrated honeycomb-lattice Ising-like antiferromagnet, respectively. We confirmed a stripy order of subsystem B below $T_{\rm N1}$. We found that the magnetic ordering of subsystem A below $T_{\rm N2}$ has the 2D nature without a long-range order along the $c$ direction. Because excitation spectra from both subsystems are well separated and independent of each other, we confirmed that these two subsystems are decoupled. We found that the excitation spectrum of subsystem A is almost the same as that of {\BCSO} when scaled by the exchange constant $J$, which indicates that the excitation spectra of subsystem A and {\BCSO} are universal in $S\,{=}\,1/2$ TLHAFs. From these findings as well as the results of calculations based on the RVB-based theory~\cite{Ferrari2019,Zhang2020}, we can deduce that the RVB state remains as a quantum fluctuation in the ordered ground state. Dispersion curves of subsystem B are well described by a $J_1{-}J_2{-}J_3$ Ising-like XXZ model on a honeycomb lattice.

\acknowledgments
This work was supported by Grants-in-Aid for Scientific Research (A) (No.~17H01142) and (C) (No.~19K03711) from Japan Society for the Promotion of Science (JSPS) and by a Grant-in-Aid for JSPS Fellows (No.~20J22215). The experiments on AMATERAS were performed with the approval of J-PARC (Proposal No. 2019B0092).


\begin{thebibliography}{46}%
\makeatletter
\providecommand \@ifxundefined [1]{%
 \@ifx{#1\undefined}
}%
\providecommand \@ifnum [1]{%
 \ifnum #1\expandafter \@firstoftwo
 \else \expandafter \@secondoftwo
 \fi
}%
\providecommand \@ifx [1]{%
 \ifx #1\expandafter \@firstoftwo
 \else \expandafter \@secondoftwo
 \fi
}%
\providecommand \natexlab [1]{#1}%
\providecommand \enquote  [1]{``#1''}%
\providecommand \bibnamefont  [1]{#1}%
\providecommand \bibfnamefont [1]{#1}%
\providecommand \citenamefont [1]{#1}%
\providecommand \href@noop [0]{\@secondoftwo}%
\providecommand \href [0]{\begingroup \@sanitize@url \@href}%
\providecommand \@href[1]{\@@startlink{#1}\@@href}%
\providecommand \@@href[1]{\endgroup#1\@@endlink}%
\providecommand \@sanitize@url [0]{\catcode `\\12\catcode `\$12\catcode
  `\&12\catcode `\#12\catcode `\^12\catcode `\_12\catcode `\%12\relax}%
\providecommand \@@startlink[1]{}%
\providecommand \@@endlink[0]{}%
\providecommand \url  [0]{\begingroup\@sanitize@url \@url }%
\providecommand \@url [1]{\endgroup\@href {#1}{\urlprefix }}%
\providecommand \urlprefix  [0]{URL }%
\providecommand \Eprint [0]{\href }%
\providecommand \doibase [0]{https://doi.org/}%
\providecommand \selectlanguage [0]{\@gobble}%
\providecommand \bibinfo  [0]{\@secondoftwo}%
\providecommand \bibfield  [0]{\@secondoftwo}%
\providecommand \translation [1]{[#1]}%
\providecommand \BibitemOpen [0]{}%
\providecommand \bibitemStop [0]{}%
\providecommand \bibitemNoStop [0]{.\EOS\space}%
\providecommand \EOS [0]{\spacefactor3000\relax}%
\providecommand \BibitemShut  [1]{\csname bibitem#1\endcsname}%
\let\auto@bib@innerbib\@empty
\bibitem [{\citenamefont {Huse}\ and\ \citenamefont {Elser}(1988)}]{Huse1988}%
  \BibitemOpen
  \bibfield  {author} {\bibinfo {author} {\bibfnamefont {D.~A.}\ \bibnamefont
  {Huse}}\ and\ \bibinfo {author} {\bibfnamefont {V.}~\bibnamefont {Elser}},\
  }\bibfield  {title} {\bibinfo {title} {Simple variational wave functions for
  two-dimensional {Heisenberg} spin-\textonehalf{} antiferromagnets},\ }\href
  {https://doi.org/10.1103/PhysRevLett.60.2531} {\bibfield  {journal} {\bibinfo
   {journal} {Phys. Rev. Lett.}\ }\textbf {\bibinfo {volume} {60}},\ \bibinfo
  {pages} {2531} (\bibinfo {year} {1988})}\BibitemShut {NoStop}%
\bibitem [{\citenamefont {Jolicoeur}\ and\ \citenamefont
  {Le~Guillou}(1989)}]{Jolicoeur1989}%
  \BibitemOpen
  \bibfield  {author} {\bibinfo {author} {\bibfnamefont {T.}~\bibnamefont
  {Jolicoeur}}\ and\ \bibinfo {author} {\bibfnamefont {J.~C.}\ \bibnamefont
  {Le~Guillou}},\ }\bibfield  {title} {\bibinfo {title} {Spin-wave results for
  the triangular {Heisenberg} antiferromagnet},\ }\href
  {https://doi.org/10.1103/PhysRevB.40.2727} {\bibfield  {journal} {\bibinfo
  {journal} {Phys. Rev. B}\ }\textbf {\bibinfo {volume} {40}},\ \bibinfo
  {pages} {2727} (\bibinfo {year} {1989})}\BibitemShut {NoStop}%
\bibitem [{\citenamefont {Bernu}\ \emph {et~al.}(1994)\citenamefont {Bernu},
  \citenamefont {Lecheminant}, \citenamefont {Lhuillier},\ and\ \citenamefont
  {Pierre}}]{Bernu1994}%
  \BibitemOpen
  \bibfield  {author} {\bibinfo {author} {\bibfnamefont {B.}~\bibnamefont
  {Bernu}}, \bibinfo {author} {\bibfnamefont {P.}~\bibnamefont {Lecheminant}},
  \bibinfo {author} {\bibfnamefont {C.}~\bibnamefont {Lhuillier}},\ and\
  \bibinfo {author} {\bibfnamefont {L.}~\bibnamefont {Pierre}},\ }\bibfield
  {title} {\bibinfo {title} {Exact spectra, spin susceptibilities, and order
  parameter of the quantum {Heisenberg} antiferromagnet on the triangular
  lattice},\ }\href {https://doi.org/10.1103/PhysRevB.50.10048} {\bibfield
  {journal} {\bibinfo  {journal} {Phys. Rev. B}\ }\textbf {\bibinfo {volume}
  {50}},\ \bibinfo {pages} {10048} (\bibinfo {year} {1994})}\BibitemShut
  {NoStop}%
\bibitem [{\citenamefont {Singh}\ and\ \citenamefont {Huse}(1992)}]{Singh1992}%
  \BibitemOpen
  \bibfield  {author} {\bibinfo {author} {\bibfnamefont {R.~R.~P.}\
  \bibnamefont {Singh}}\ and\ \bibinfo {author} {\bibfnamefont {D.~A.}\
  \bibnamefont {Huse}},\ }\bibfield  {title} {\bibinfo {title}
  {Three-sublattice order in triangular- and {Kagom\'e}-lattice spin-half
  antiferromagnets},\ }\href {https://doi.org/10.1103/PhysRevLett.68.1766}
  {\bibfield  {journal} {\bibinfo  {journal} {Phys. Rev. Lett.}\ }\textbf
  {\bibinfo {volume} {68}},\ \bibinfo {pages} {1766} (\bibinfo {year}
  {1992})}\BibitemShut {NoStop}%
\bibitem [{\citenamefont {White}\ and\ \citenamefont
  {Chernyshev}(2007)}]{White2007}%
  \BibitemOpen
  \bibfield  {author} {\bibinfo {author} {\bibfnamefont {S.~R.}\ \bibnamefont
  {White}}\ and\ \bibinfo {author} {\bibfnamefont {A.~L.}\ \bibnamefont
  {Chernyshev}},\ }\bibfield  {title} {\bibinfo {title} {{Ne\'el} order in
  square and triangular lattice {Heisenberg} models},\ }\href
  {https://doi.org/10.1103/PhysRevLett.99.127004} {\bibfield  {journal}
  {\bibinfo  {journal} {Phys. Rev. Lett.}\ }\textbf {\bibinfo {volume} {99}},\
  \bibinfo {pages} {127004} (\bibinfo {year} {2007})}\BibitemShut {NoStop}%
\bibitem [{\citenamefont {Götze}\ \emph {et~al.}(2016)\citenamefont {Götze},
  \citenamefont {Richter}, \citenamefont {Zinke},\ and\ \citenamefont
  {Farnell}}]{Gotze2016}%
  \BibitemOpen
  \bibfield  {author} {\bibinfo {author} {\bibfnamefont {O.}~\bibnamefont
  {Götze}}, \bibinfo {author} {\bibfnamefont {J.}~\bibnamefont {Richter}},
  \bibinfo {author} {\bibfnamefont {R.}~\bibnamefont {Zinke}},\ and\ \bibinfo
  {author} {\bibfnamefont {D.}~\bibnamefont {Farnell}},\ }\bibfield  {title}
  {\bibinfo {title} {Ground-state properties of the triangular-lattice
  {Heisenberg} antiferromagnet with arbitrary spin quantum number $s$},\ }\href
  {https://doi.org/https://doi.org/10.1016/j.jmmm.2015.08.113} {\bibfield
  {journal} {\bibinfo  {journal} {J. Magn. Magn. Mater.}\ }\textbf {\bibinfo
  {volume} {397}},\ \bibinfo {pages} {333} (\bibinfo {year}
  {2016})}\BibitemShut {NoStop}%
\bibitem [{\citenamefont {Anderson}(1973)}]{Anderson1973}%
  \BibitemOpen
  \bibfield  {author} {\bibinfo {author} {\bibfnamefont {P.}~\bibnamefont
  {Anderson}},\ }\bibfield  {title} {\bibinfo {title} {Resonating valence
  bonds: A new kind of insulator?},\ }\href
  {https://doi.org/https://doi.org/10.1016/0025-5408(73)90167-0} {\bibfield
  {journal} {\bibinfo  {journal} {Mater. Res. Bull.}\ }\textbf {\bibinfo
  {volume} {8}},\ \bibinfo {pages} {153} (\bibinfo {year} {1973})}\BibitemShut
  {NoStop}%
\bibitem [{\citenamefont {Chubukov}\ and\ \citenamefont
  {Golosov}(1991)}]{Chubukov1991}%
  \BibitemOpen
  \bibfield  {author} {\bibinfo {author} {\bibfnamefont {A.~V.}\ \bibnamefont
  {Chubukov}}\ and\ \bibinfo {author} {\bibfnamefont {D.~I.}\ \bibnamefont
  {Golosov}},\ }\bibfield  {title} {\bibinfo {title} {Quantum theory of an
  antiferromagnet on a triangular lattice in a magnetic field},\ }\href
  {https://doi.org/10.1088/0953-8984/3/1/005} {\bibfield  {journal} {\bibinfo
  {journal} {J. Phys.: Condens. Matter}\ }\textbf {\bibinfo {volume} {3}},\
  \bibinfo {pages} {69} (\bibinfo {year} {1991})}\BibitemShut {NoStop}%
\bibitem [{\citenamefont {Nikuni}\ and\ \citenamefont
  {Shiba}(1993)}]{Nikuni1993}%
  \BibitemOpen
  \bibfield  {author} {\bibinfo {author} {\bibfnamefont {T.}~\bibnamefont
  {Nikuni}}\ and\ \bibinfo {author} {\bibfnamefont {H.}~\bibnamefont {Shiba}},\
  }\bibfield  {title} {\bibinfo {title} {Quantum fluctuations and magnetic
  structures of {$\mathrm{CsCuCl}_3$} in high magnetic field},\ }\href
  {https://doi.org/10.1143/JPSJ.62.3268} {\bibfield  {journal} {\bibinfo
  {journal} {J. Phys. Soc. Jpn.}\ }\textbf {\bibinfo {volume} {62}},\ \bibinfo
  {pages} {3268} (\bibinfo {year} {1993})}\BibitemShut {NoStop}%
\bibitem [{\citenamefont {Honecker}(1999)}]{Honecker1999}%
  \BibitemOpen
  \bibfield  {author} {\bibinfo {author} {\bibfnamefont {A.}~\bibnamefont
  {Honecker}},\ }\bibfield  {title} {\bibinfo {title} {A comparative study of
  the magnetization process of two-dimensional antiferromagnets},\ }\href
  {https://doi.org/10.1088/0953-8984/11/24/311} {\bibfield  {journal} {\bibinfo
   {journal} {J. Phys.: Condens. Matter}\ }\textbf {\bibinfo {volume} {11}},\
  \bibinfo {pages} {4697} (\bibinfo {year} {1999})}\BibitemShut {NoStop}%
\bibitem [{\citenamefont {Alicea}\ \emph {et~al.}(2009)\citenamefont {Alicea},
  \citenamefont {Chubukov},\ and\ \citenamefont {Starykh}}]{Alicea2009}%
  \BibitemOpen
  \bibfield  {author} {\bibinfo {author} {\bibfnamefont {J.}~\bibnamefont
  {Alicea}}, \bibinfo {author} {\bibfnamefont {A.~V.}\ \bibnamefont
  {Chubukov}},\ and\ \bibinfo {author} {\bibfnamefont {O.~A.}\ \bibnamefont
  {Starykh}},\ }\bibfield  {title} {\bibinfo {title} {Quantum stabilization of
  the $1/3$-magnetization plateau in
  {${\mathrm{Cs}}_{2}{\mathrm{CuBr}}_{4}$}},\ }\href
  {https://doi.org/10.1103/PhysRevLett.102.137201} {\bibfield  {journal}
  {\bibinfo  {journal} {Phys. Rev. Lett.}\ }\textbf {\bibinfo {volume} {102}},\
  \bibinfo {pages} {137201} (\bibinfo {year} {2009})}\BibitemShut {NoStop}%
\bibitem [{\citenamefont {Farnell}\ \emph {et~al.}(2009)\citenamefont
  {Farnell}, \citenamefont {Zinke}, \citenamefont {Schulenburg},\ and\
  \citenamefont {Richter}}]{Farnell2009}%
  \BibitemOpen
  \bibfield  {author} {\bibinfo {author} {\bibfnamefont {D.~J.~J.}\
  \bibnamefont {Farnell}}, \bibinfo {author} {\bibfnamefont {R.}~\bibnamefont
  {Zinke}}, \bibinfo {author} {\bibfnamefont {J.}~\bibnamefont {Schulenburg}},\
  and\ \bibinfo {author} {\bibfnamefont {J.}~\bibnamefont {Richter}},\
  }\bibfield  {title} {\bibinfo {title} {High-order coupled cluster method
  study of frustrated and unfrustrated quantum magnets in external magnetic
  fields},\ }\href {https://doi.org/10.1088/0953-8984/21/40/406002} {\bibfield
  {journal} {\bibinfo  {journal} {J. Phys.: Condens. Matter}\ }\textbf
  {\bibinfo {volume} {21}},\ \bibinfo {pages} {406002} (\bibinfo {year}
  {2009})}\BibitemShut {NoStop}%
\bibitem [{\citenamefont {Sakai}\ and\ \citenamefont
  {Nakano}(2011)}]{Sakai2011}%
  \BibitemOpen
  \bibfield  {author} {\bibinfo {author} {\bibfnamefont {T.}~\bibnamefont
  {Sakai}}\ and\ \bibinfo {author} {\bibfnamefont {H.}~\bibnamefont {Nakano}},\
  }\bibfield  {title} {\bibinfo {title} {Critical magnetization behavior of the
  triangular- and kagome-lattice quantum antiferromagnets},\ }\href
  {https://doi.org/10.1103/PhysRevB.83.100405} {\bibfield  {journal} {\bibinfo
  {journal} {Phys. Rev. B}\ }\textbf {\bibinfo {volume} {83}},\ \bibinfo
  {pages} {100405(R)} (\bibinfo {year} {2011})}\BibitemShut {NoStop}%
\bibitem [{\citenamefont {Hotta}\ \emph {et~al.}(2013)\citenamefont {Hotta},
  \citenamefont {Nishimoto},\ and\ \citenamefont {Shibata}}]{Hotta2013}%
  \BibitemOpen
  \bibfield  {author} {\bibinfo {author} {\bibfnamefont {C.}~\bibnamefont
  {Hotta}}, \bibinfo {author} {\bibfnamefont {S.}~\bibnamefont {Nishimoto}},\
  and\ \bibinfo {author} {\bibfnamefont {N.}~\bibnamefont {Shibata}},\
  }\bibfield  {title} {\bibinfo {title} {Grand canonical finite size numerical
  approaches in one and two dimensions: Real space energy renormalization and
  edge state generation},\ }\href {https://doi.org/10.1103/PhysRevB.87.115128}
  {\bibfield  {journal} {\bibinfo  {journal} {Phys. Rev. B}\ }\textbf {\bibinfo
  {volume} {87}},\ \bibinfo {pages} {115128} (\bibinfo {year}
  {2013})}\BibitemShut {NoStop}%
\bibitem [{\citenamefont {Yamamoto}\ \emph {et~al.}(2014)\citenamefont
  {Yamamoto}, \citenamefont {Marmorini},\ and\ \citenamefont
  {Danshita}}]{Yamamoto2014}%
  \BibitemOpen
  \bibfield  {author} {\bibinfo {author} {\bibfnamefont {D.}~\bibnamefont
  {Yamamoto}}, \bibinfo {author} {\bibfnamefont {G.}~\bibnamefont
  {Marmorini}},\ and\ \bibinfo {author} {\bibfnamefont {I.}~\bibnamefont
  {Danshita}},\ }\bibfield  {title} {\bibinfo {title} {Quantum phase diagram of
  the triangular-lattice {$XXZ$} model in a magnetic field},\ }\href
  {https://doi.org/10.1103/PhysRevLett.112.127203} {\bibfield  {journal}
  {\bibinfo  {journal} {Phys. Rev. Lett.}\ }\textbf {\bibinfo {volume} {112}},\
  \bibinfo {pages} {127203} (\bibinfo {year} {2014})}\BibitemShut {NoStop}%
\bibitem [{\citenamefont {Starykh}(2015)}]{Starykh2015}%
  \BibitemOpen
  \bibfield  {author} {\bibinfo {author} {\bibfnamefont {O.~A.}\ \bibnamefont
  {Starykh}},\ }\bibfield  {title} {\bibinfo {title} {Unusual ordered phases of
  highly frustrated magnets: a review},\ }\href
  {https://doi.org/10.1088/0034-4885/78/5/052502} {\bibfield  {journal}
  {\bibinfo  {journal} {Rep. Prog. Phys.}\ }\textbf {\bibinfo {volume} {78}},\
  \bibinfo {pages} {052502} (\bibinfo {year} {2015})}\BibitemShut {NoStop}%
\bibitem [{\citenamefont {Sellmann}\ \emph {et~al.}(2015)\citenamefont
  {Sellmann}, \citenamefont {Zhang},\ and\ \citenamefont
  {Eggert}}]{Sellmann2015}%
  \BibitemOpen
  \bibfield  {author} {\bibinfo {author} {\bibfnamefont {D.}~\bibnamefont
  {Sellmann}}, \bibinfo {author} {\bibfnamefont {X.-F.}\ \bibnamefont
  {Zhang}},\ and\ \bibinfo {author} {\bibfnamefont {S.}~\bibnamefont
  {Eggert}},\ }\bibfield  {title} {\bibinfo {title} {Phase diagram of the
  antiferromagnetic {XXZ} model on the triangular lattice},\ }\href
  {https://doi.org/10.1103/PhysRevB.91.081104} {\bibfield  {journal} {\bibinfo
  {journal} {Phys. Rev. B}\ }\textbf {\bibinfo {volume} {91}},\ \bibinfo
  {pages} {081104(R)} (\bibinfo {year} {2015})}\BibitemShut {NoStop}%
\bibitem [{\citenamefont {Coletta}\ \emph {et~al.}(2016)\citenamefont
  {Coletta}, \citenamefont {T\'oth}, \citenamefont {Penc},\ and\ \citenamefont
  {Mila}}]{Coletta2016}%
  \BibitemOpen
  \bibfield  {author} {\bibinfo {author} {\bibfnamefont {T.}~\bibnamefont
  {Coletta}}, \bibinfo {author} {\bibfnamefont {T.~A.}\ \bibnamefont {T\'oth}},
  \bibinfo {author} {\bibfnamefont {K.}~\bibnamefont {Penc}},\ and\ \bibinfo
  {author} {\bibfnamefont {F.}~\bibnamefont {Mila}},\ }\bibfield  {title}
  {\bibinfo {title} {Semiclassical theory of the magnetization process of the
  triangular lattice {Heisenberg} model},\ }\href
  {https://doi.org/10.1103/PhysRevB.94.075136} {\bibfield  {journal} {\bibinfo
  {journal} {Phys. Rev. B}\ }\textbf {\bibinfo {volume} {94}},\ \bibinfo
  {pages} {075136} (\bibinfo {year} {2016})}\BibitemShut {NoStop}%
\bibitem [{\citenamefont {Starykh}\ \emph {et~al.}(2006)\citenamefont
  {Starykh}, \citenamefont {Chubukov},\ and\ \citenamefont
  {Abanov}}]{Starykh2006}%
  \BibitemOpen
  \bibfield  {author} {\bibinfo {author} {\bibfnamefont {O.~A.}\ \bibnamefont
  {Starykh}}, \bibinfo {author} {\bibfnamefont {A.~V.}\ \bibnamefont
  {Chubukov}},\ and\ \bibinfo {author} {\bibfnamefont {A.~G.}\ \bibnamefont
  {Abanov}},\ }\bibfield  {title} {\bibinfo {title} {Flat spin-wave dispersion
  in a triangular antiferromagnet},\ }\href
  {https://doi.org/10.1103/PhysRevB.74.180403} {\bibfield  {journal} {\bibinfo
  {journal} {Phys. Rev. B}\ }\textbf {\bibinfo {volume} {74}},\ \bibinfo
  {pages} {180403(R)} (\bibinfo {year} {2006})}\BibitemShut {NoStop}%
\bibitem [{\citenamefont {Zheng}\ \emph {et~al.}(2006)\citenamefont {Zheng},
  \citenamefont {Fj\ae{}restad}, \citenamefont {Singh}, \citenamefont
  {McKenzie},\ and\ \citenamefont {Coldea}}]{Zheng2006}%
  \BibitemOpen
  \bibfield  {author} {\bibinfo {author} {\bibfnamefont {W.}~\bibnamefont
  {Zheng}}, \bibinfo {author} {\bibfnamefont {J.~O.}\ \bibnamefont
  {Fj\ae{}restad}}, \bibinfo {author} {\bibfnamefont {R.~R.~P.}\ \bibnamefont
  {Singh}}, \bibinfo {author} {\bibfnamefont {R.~H.}\ \bibnamefont
  {McKenzie}},\ and\ \bibinfo {author} {\bibfnamefont {R.}~\bibnamefont
  {Coldea}},\ }\bibfield  {title} {\bibinfo {title} {Excitation spectra of the
  spin-$\frac{1}{2}$ triangular-lattice {Heisenberg} antiferromagnet},\ }\href
  {https://doi.org/10.1103/PhysRevB.74.224420} {\bibfield  {journal} {\bibinfo
  {journal} {Phys. Rev. B}\ }\textbf {\bibinfo {volume} {74}},\ \bibinfo
  {pages} {224420} (\bibinfo {year} {2006})}\BibitemShut {NoStop}%
\bibitem [{\citenamefont {Chernyshev}\ and\ \citenamefont
  {Zhitomirsky}(2009)}]{Chernyshev2009}%
  \BibitemOpen
  \bibfield  {author} {\bibinfo {author} {\bibfnamefont {A.~L.}\ \bibnamefont
  {Chernyshev}}\ and\ \bibinfo {author} {\bibfnamefont {M.~E.}\ \bibnamefont
  {Zhitomirsky}},\ }\bibfield  {title} {\bibinfo {title} {Spin waves in a
  triangular lattice antiferromagnet: Decays, spectrum renormalization, and
  singularities},\ }\href {https://doi.org/10.1103/PhysRevB.79.144416}
  {\bibfield  {journal} {\bibinfo  {journal} {Phys. Rev. B}\ }\textbf {\bibinfo
  {volume} {79}},\ \bibinfo {pages} {144416} (\bibinfo {year}
  {2009})}\BibitemShut {NoStop}%
\bibitem [{\citenamefont {Mezio}\ \emph {et~al.}(2011)\citenamefont {Mezio},
  \citenamefont {Sposetti}, \citenamefont {Manuel},\ and\ \citenamefont
  {Trumper}}]{Mezio2011}%
  \BibitemOpen
  \bibfield  {author} {\bibinfo {author} {\bibfnamefont {A.}~\bibnamefont
  {Mezio}}, \bibinfo {author} {\bibfnamefont {C.~N.}\ \bibnamefont {Sposetti}},
  \bibinfo {author} {\bibfnamefont {L.~O.}\ \bibnamefont {Manuel}},\ and\
  \bibinfo {author} {\bibfnamefont {A.~E.}\ \bibnamefont {Trumper}},\
  }\bibfield  {title} {\bibinfo {title} {A test of the bosonic spinon theory
  for the triangular antiferromagnet spectrum},\ }\href
  {https://doi.org/10.1209/0295-5075/94/47001} {\bibfield  {journal} {\bibinfo
  {journal} {Europhys. Lett.}\ }\textbf {\bibinfo {volume} {94}},\ \bibinfo
  {pages} {47001} (\bibinfo {year} {2011})}\BibitemShut {NoStop}%
\bibitem [{\citenamefont {Mourigal}\ \emph {et~al.}(2013)\citenamefont
  {Mourigal}, \citenamefont {Fuhrman}, \citenamefont {Chernyshev},\ and\
  \citenamefont {Zhitomirsky}}]{Mourigal2013}%
  \BibitemOpen
  \bibfield  {author} {\bibinfo {author} {\bibfnamefont {M.}~\bibnamefont
  {Mourigal}}, \bibinfo {author} {\bibfnamefont {W.~T.}\ \bibnamefont
  {Fuhrman}}, \bibinfo {author} {\bibfnamefont {A.~L.}\ \bibnamefont
  {Chernyshev}},\ and\ \bibinfo {author} {\bibfnamefont {M.~E.}\ \bibnamefont
  {Zhitomirsky}},\ }\bibfield  {title} {\bibinfo {title} {Dynamical structure
  factor of the triangular-lattice antiferromagnet},\ }\href
  {https://doi.org/10.1103/PhysRevB.88.094407} {\bibfield  {journal} {\bibinfo
  {journal} {Phys. Rev. B}\ }\textbf {\bibinfo {volume} {88}},\ \bibinfo
  {pages} {094407} (\bibinfo {year} {2013})}\BibitemShut {NoStop}%
\bibitem [{\citenamefont {Ghioldi}\ \emph {et~al.}(2015)\citenamefont
  {Ghioldi}, \citenamefont {Mezio}, \citenamefont {Manuel}, \citenamefont
  {Singh}, \citenamefont {Oitmaa},\ and\ \citenamefont
  {Trumper}}]{Ghioldi2015}%
  \BibitemOpen
  \bibfield  {author} {\bibinfo {author} {\bibfnamefont {E.~A.}\ \bibnamefont
  {Ghioldi}}, \bibinfo {author} {\bibfnamefont {A.}~\bibnamefont {Mezio}},
  \bibinfo {author} {\bibfnamefont {L.~O.}\ \bibnamefont {Manuel}}, \bibinfo
  {author} {\bibfnamefont {R.~R.~P.}\ \bibnamefont {Singh}}, \bibinfo {author}
  {\bibfnamefont {J.}~\bibnamefont {Oitmaa}},\ and\ \bibinfo {author}
  {\bibfnamefont {A.~E.}\ \bibnamefont {Trumper}},\ }\bibfield  {title}
  {\bibinfo {title} {Magnons and excitation continuum in {XXZ} triangular
  antiferromagnetic model: Application to
  {${\mathrm{Ba}}_{3}{\mathrm{CoSb}}_{2}{\mathrm{O}}_{9}$}},\ }\href
  {https://doi.org/10.1103/PhysRevB.91.134423} {\bibfield  {journal} {\bibinfo
  {journal} {Phys. Rev. B}\ }\textbf {\bibinfo {volume} {91}},\ \bibinfo
  {pages} {134423} (\bibinfo {year} {2015})}\BibitemShut {NoStop}%
\bibitem [{\citenamefont {Ma}\ \emph {et~al.}(2016)\citenamefont {Ma},
  \citenamefont {Kamiya}, \citenamefont {Hong}, \citenamefont {Cao},
  \citenamefont {Ehlers}, \citenamefont {Tian}, \citenamefont {Batista},
  \citenamefont {Dun}, \citenamefont {Zhou},\ and\ \citenamefont
  {Matsuda}}]{Ma2016}%
  \BibitemOpen
  \bibfield  {author} {\bibinfo {author} {\bibfnamefont {J.}~\bibnamefont
  {Ma}}, \bibinfo {author} {\bibfnamefont {Y.}~\bibnamefont {Kamiya}}, \bibinfo
  {author} {\bibfnamefont {T.}~\bibnamefont {Hong}}, \bibinfo {author}
  {\bibfnamefont {H.~B.}\ \bibnamefont {Cao}}, \bibinfo {author} {\bibfnamefont
  {G.}~\bibnamefont {Ehlers}}, \bibinfo {author} {\bibfnamefont
  {W.}~\bibnamefont {Tian}}, \bibinfo {author} {\bibfnamefont {C.~D.}\
  \bibnamefont {Batista}}, \bibinfo {author} {\bibfnamefont {Z.~L.}\
  \bibnamefont {Dun}}, \bibinfo {author} {\bibfnamefont {H.~D.}\ \bibnamefont
  {Zhou}},\ and\ \bibinfo {author} {\bibfnamefont {M.}~\bibnamefont
  {Matsuda}},\ }\bibfield  {title} {\bibinfo {title} {Static and dynamical
  properties of the spin-$1/2$ equilateral triangular-lattice antiferromagnet
  {${\mathrm{Ba}}_{3}{\mathrm{CoSb}}_{2}{\mathrm{O}}_{9}$}},\ }\href
  {https://doi.org/10.1103/PhysRevLett.116.087201} {\bibfield  {journal}
  {\bibinfo  {journal} {Phys. Rev. Lett.}\ }\textbf {\bibinfo {volume} {116}},\
  \bibinfo {pages} {087201} (\bibinfo {year} {2016})}\BibitemShut {NoStop}%
\bibitem [{\citenamefont {Ito}\ \emph {et~al.}(2017)\citenamefont {Ito},
  \citenamefont {Kurita}, \citenamefont {Tanaka}, \citenamefont
  {Ohira-Kawamura}, \citenamefont {Nakajima}, \citenamefont {Itoh},
  \citenamefont {Kuwahara},\ and\ \citenamefont {Kakurai}}]{Ito2017}%
  \BibitemOpen
  \bibfield  {author} {\bibinfo {author} {\bibfnamefont {S.}~\bibnamefont
  {Ito}}, \bibinfo {author} {\bibfnamefont {N.}~\bibnamefont {Kurita}},
  \bibinfo {author} {\bibfnamefont {H.}~\bibnamefont {Tanaka}}, \bibinfo
  {author} {\bibfnamefont {S.}~\bibnamefont {Ohira-Kawamura}}, \bibinfo
  {author} {\bibfnamefont {K.}~\bibnamefont {Nakajima}}, \bibinfo {author}
  {\bibfnamefont {S.}~\bibnamefont {Itoh}}, \bibinfo {author} {\bibfnamefont
  {K.}~\bibnamefont {Kuwahara}},\ and\ \bibinfo {author} {\bibfnamefont
  {K.}~\bibnamefont {Kakurai}},\ }\bibfield  {title} {\bibinfo {title}
  {Structure of the magnetic excitations in the spin-$1/2$ triangular-lattice
  {Heisenberg} antiferromagnet
  {${\mathrm{Ba}}_{3}{\mathrm{CoSb}}_{2}{\mathrm{O}}_{9}$}},\ }\href
  {https://doi.org/10.1038/s41467-017-00316-x} {\bibfield  {journal} {\bibinfo
  {journal} {Nat. Comm.}\ }\textbf {\bibinfo {volume} {8}},\ \bibinfo {pages}
  {235} (\bibinfo {year} {2017})}\BibitemShut {NoStop}%
\bibitem [{\citenamefont {Kamiya}\ \emph {et~al.}(2018)\citenamefont {Kamiya},
  \citenamefont {Ge}, \citenamefont {Hong}, \citenamefont {Qiu}, \citenamefont
  {Quintero-Castro}, \citenamefont {Lu}, \citenamefont {Cao}, \citenamefont
  {Matsuda}, \citenamefont {Choi}, \citenamefont {Batista}, \citenamefont
  {Mourigal}, \citenamefont {Zhou},\ and\ \citenamefont {Ma}}]{Kamiya2018}%
  \BibitemOpen
  \bibfield  {author} {\bibinfo {author} {\bibfnamefont {Y.}~\bibnamefont
  {Kamiya}}, \bibinfo {author} {\bibfnamefont {L.}~\bibnamefont {Ge}}, \bibinfo
  {author} {\bibfnamefont {T.}~\bibnamefont {Hong}}, \bibinfo {author}
  {\bibfnamefont {Y.}~\bibnamefont {Qiu}}, \bibinfo {author} {\bibfnamefont
  {D.~L.}\ \bibnamefont {Quintero-Castro}}, \bibinfo {author} {\bibfnamefont
  {Z.}~\bibnamefont {Lu}}, \bibinfo {author} {\bibfnamefont {H.~B.}\
  \bibnamefont {Cao}}, \bibinfo {author} {\bibfnamefont {M.}~\bibnamefont
  {Matsuda}}, \bibinfo {author} {\bibfnamefont {E.~S.}\ \bibnamefont {Choi}},
  \bibinfo {author} {\bibfnamefont {C.~D.}\ \bibnamefont {Batista}}, \bibinfo
  {author} {\bibfnamefont {M.}~\bibnamefont {Mourigal}}, \bibinfo {author}
  {\bibfnamefont {H.~D.}\ \bibnamefont {Zhou}},\ and\ \bibinfo {author}
  {\bibfnamefont {J.}~\bibnamefont {Ma}},\ }\bibfield  {title} {\bibinfo
  {title} {The nature of spin excitations in the one-third magnetization
  plateau phase of {${\mathrm{Ba}}_{3}{\mathrm{CoSb}}_{2}{\mathrm{O}}_{9}$}},\
  }\href {https://doi.org/10.1038/s41467-018-04914-1} {\bibfield  {journal}
  {\bibinfo  {journal} {Nat. Comm.}\ }\textbf {\bibinfo {volume} {9}},\
  \bibinfo {pages} {2666} (\bibinfo {year} {2018})}\BibitemShut {NoStop}%
\bibitem [{\citenamefont {Macdougal}\ \emph {et~al.}(2020)\citenamefont
  {Macdougal}, \citenamefont {Williams}, \citenamefont {Prabhakaran},
  \citenamefont {Bewley}, \citenamefont {Voneshen},\ and\ \citenamefont
  {Coldea}}]{Macdougal2020}%
  \BibitemOpen
  \bibfield  {author} {\bibinfo {author} {\bibfnamefont {D.}~\bibnamefont
  {Macdougal}}, \bibinfo {author} {\bibfnamefont {S.}~\bibnamefont {Williams}},
  \bibinfo {author} {\bibfnamefont {D.}~\bibnamefont {Prabhakaran}}, \bibinfo
  {author} {\bibfnamefont {R.~I.}\ \bibnamefont {Bewley}}, \bibinfo {author}
  {\bibfnamefont {D.~J.}\ \bibnamefont {Voneshen}},\ and\ \bibinfo {author}
  {\bibfnamefont {R.}~\bibnamefont {Coldea}},\ }\bibfield  {title} {\bibinfo
  {title} {Avoided quasiparticle decay and enhanced excitation continuum in the
  spin-$\frac{1}{2}$ near-{Heisenberg} triangular antiferromagnet
  {${\mathrm{Ba}}_{3}{\mathrm{CoSb}}_{2}{\mathrm{O}}_{9}$}},\ }\href
  {https://doi.org/10.1103/PhysRevB.102.064421} {\bibfield  {journal} {\bibinfo
   {journal} {Phys. Rev. B}\ }\textbf {\bibinfo {volume} {102}},\ \bibinfo
  {pages} {064421} (\bibinfo {year} {2020})}\BibitemShut {NoStop}%
\bibitem [{\citenamefont {Shirata}\ \emph {et~al.}(2012)\citenamefont
  {Shirata}, \citenamefont {Tanaka}, \citenamefont {Matsuo},\ and\
  \citenamefont {Kindo}}]{Shirata2012}%
  \BibitemOpen
  \bibfield  {author} {\bibinfo {author} {\bibfnamefont {Y.}~\bibnamefont
  {Shirata}}, \bibinfo {author} {\bibfnamefont {H.}~\bibnamefont {Tanaka}},
  \bibinfo {author} {\bibfnamefont {A.}~\bibnamefont {Matsuo}},\ and\ \bibinfo
  {author} {\bibfnamefont {K.}~\bibnamefont {Kindo}},\ }\bibfield  {title}
  {\bibinfo {title} {Experimental realization of a spin-$1/2$
  triangular-lattice {Heisenberg} antiferromagnet},\ }\href
  {https://doi.org/10.1103/PhysRevLett.108.057205} {\bibfield  {journal}
  {\bibinfo  {journal} {Phys. Rev. Lett.}\ }\textbf {\bibinfo {volume} {108}},\
  \bibinfo {pages} {057205} (\bibinfo {year} {2012})}\BibitemShut {NoStop}%
\bibitem [{\citenamefont {Zhou}\ \emph {et~al.}(2012)\citenamefont {Zhou},
  \citenamefont {Xu}, \citenamefont {Hallas}, \citenamefont {Silverstein},
  \citenamefont {Wiebe}, \citenamefont {Umegaki}, \citenamefont {Yan},
  \citenamefont {Murphy}, \citenamefont {Park}, \citenamefont {Qiu},
  \citenamefont {Copley}, \citenamefont {Gardner},\ and\ \citenamefont
  {Takano}}]{Zhou2012}%
  \BibitemOpen
  \bibfield  {author} {\bibinfo {author} {\bibfnamefont {H.~D.}\ \bibnamefont
  {Zhou}}, \bibinfo {author} {\bibfnamefont {C.}~\bibnamefont {Xu}}, \bibinfo
  {author} {\bibfnamefont {A.~M.}\ \bibnamefont {Hallas}}, \bibinfo {author}
  {\bibfnamefont {H.~J.}\ \bibnamefont {Silverstein}}, \bibinfo {author}
  {\bibfnamefont {C.~R.}\ \bibnamefont {Wiebe}}, \bibinfo {author}
  {\bibfnamefont {I.}~\bibnamefont {Umegaki}}, \bibinfo {author} {\bibfnamefont
  {J.~Q.}\ \bibnamefont {Yan}}, \bibinfo {author} {\bibfnamefont {T.~P.}\
  \bibnamefont {Murphy}}, \bibinfo {author} {\bibfnamefont {J.-H.}\
  \bibnamefont {Park}}, \bibinfo {author} {\bibfnamefont {Y.}~\bibnamefont
  {Qiu}}, \bibinfo {author} {\bibfnamefont {J.~R.~D.}\ \bibnamefont {Copley}},
  \bibinfo {author} {\bibfnamefont {J.~S.}\ \bibnamefont {Gardner}},\ and\
  \bibinfo {author} {\bibfnamefont {Y.}~\bibnamefont {Takano}},\ }\bibfield
  {title} {\bibinfo {title} {Successive phase transitions and extended
  spin-excitation continuum in the {$S\mathbf{=}\frac{1}{2}$}
  triangular-lattice antiferromagnet
  {${\mathrm{Ba}}_{3}{\mathrm{CoSb}}_{2}{\mathrm{O}}_{9}$}},\ }\href
  {https://doi.org/10.1103/PhysRevLett.109.267206} {\bibfield  {journal}
  {\bibinfo  {journal} {Phys. Rev. Lett.}\ }\textbf {\bibinfo {volume} {109}},\
  \bibinfo {pages} {267206} (\bibinfo {year} {2012})}\BibitemShut {NoStop}%
\bibitem [{\citenamefont {Susuki}\ \emph {et~al.}(2013)\citenamefont {Susuki},
  \citenamefont {Kurita}, \citenamefont {Tanaka}, \citenamefont {Nojiri},
  \citenamefont {Matsuo}, \citenamefont {Kindo},\ and\ \citenamefont
  {Tanaka}}]{Susuki2013}%
  \BibitemOpen
  \bibfield  {author} {\bibinfo {author} {\bibfnamefont {T.}~\bibnamefont
  {Susuki}}, \bibinfo {author} {\bibfnamefont {N.}~\bibnamefont {Kurita}},
  \bibinfo {author} {\bibfnamefont {T.}~\bibnamefont {Tanaka}}, \bibinfo
  {author} {\bibfnamefont {H.}~\bibnamefont {Nojiri}}, \bibinfo {author}
  {\bibfnamefont {A.}~\bibnamefont {Matsuo}}, \bibinfo {author} {\bibfnamefont
  {K.}~\bibnamefont {Kindo}},\ and\ \bibinfo {author} {\bibfnamefont
  {H.}~\bibnamefont {Tanaka}},\ }\bibfield  {title} {\bibinfo {title}
  {Magnetization process and collective excitations in the {$S\mathrm{=}1/2$}
  triangular-lattice {Heisenberg} antiferromagnet
  {${\mathrm{Ba}}_{3}{\mathrm{CoSb}}_{2}{\mathrm{O}}_{9}$}},\ }\href
  {https://doi.org/10.1103/PhysRevLett.110.267201} {\bibfield  {journal}
  {\bibinfo  {journal} {Phys. Rev. Lett.}\ }\textbf {\bibinfo {volume} {110}},\
  \bibinfo {pages} {267201} (\bibinfo {year} {2013})}\BibitemShut {NoStop}%
\bibitem [{\citenamefont {Quirion}\ \emph {et~al.}(2015)\citenamefont
  {Quirion}, \citenamefont {Lapointe-Major}, \citenamefont {Poirier},
  \citenamefont {Quilliam}, \citenamefont {Dun},\ and\ \citenamefont
  {Zhou}}]{Quirion2015}%
  \BibitemOpen
  \bibfield  {author} {\bibinfo {author} {\bibfnamefont {G.}~\bibnamefont
  {Quirion}}, \bibinfo {author} {\bibfnamefont {M.}~\bibnamefont
  {Lapointe-Major}}, \bibinfo {author} {\bibfnamefont {M.}~\bibnamefont
  {Poirier}}, \bibinfo {author} {\bibfnamefont {J.~A.}\ \bibnamefont
  {Quilliam}}, \bibinfo {author} {\bibfnamefont {Z.~L.}\ \bibnamefont {Dun}},\
  and\ \bibinfo {author} {\bibfnamefont {H.~D.}\ \bibnamefont {Zhou}},\
  }\bibfield  {title} {\bibinfo {title} {Magnetic phase diagram of
  {${\mathrm{Ba}}_{3}{\mathrm{CoSb}}_{2}{\mathrm{O}}_{9}$} as determined by
  ultrasound velocity measurements},\ }\href
  {https://doi.org/10.1103/PhysRevB.92.014414} {\bibfield  {journal} {\bibinfo
  {journal} {Phys. Rev. B}\ }\textbf {\bibinfo {volume} {92}},\ \bibinfo
  {pages} {014414} (\bibinfo {year} {2015})}\BibitemShut {NoStop}%
\bibitem [{\citenamefont {Koutroulakis}\ \emph {et~al.}(2015)\citenamefont
  {Koutroulakis}, \citenamefont {Zhou}, \citenamefont {Kamiya}, \citenamefont
  {Thompson}, \citenamefont {Zhou}, \citenamefont {Batista},\ and\
  \citenamefont {Brown}}]{Koutroulakis2015}%
  \BibitemOpen
  \bibfield  {author} {\bibinfo {author} {\bibfnamefont {G.}~\bibnamefont
  {Koutroulakis}}, \bibinfo {author} {\bibfnamefont {T.}~\bibnamefont {Zhou}},
  \bibinfo {author} {\bibfnamefont {Y.}~\bibnamefont {Kamiya}}, \bibinfo
  {author} {\bibfnamefont {J.~D.}\ \bibnamefont {Thompson}}, \bibinfo {author}
  {\bibfnamefont {H.~D.}\ \bibnamefont {Zhou}}, \bibinfo {author}
  {\bibfnamefont {C.~D.}\ \bibnamefont {Batista}},\ and\ \bibinfo {author}
  {\bibfnamefont {S.~E.}\ \bibnamefont {Brown}},\ }\bibfield  {title} {\bibinfo
  {title} {Quantum phase diagram of the {$S=\frac{1}{2}$} triangular-lattice
  antiferromagnet {${\mathrm{Ba}}_{3}{\mathrm{CoSb}}_{2}{\mathrm{O}}_{9}$}},\
  }\href {https://doi.org/10.1103/PhysRevB.91.024410} {\bibfield  {journal}
  {\bibinfo  {journal} {Phys. Rev. B}\ }\textbf {\bibinfo {volume} {91}},\
  \bibinfo {pages} {024410} (\bibinfo {year} {2015})}\BibitemShut {NoStop}%
\bibitem [{\citenamefont {Yamamoto}\ \emph {et~al.}(2015)\citenamefont
  {Yamamoto}, \citenamefont {Marmorini},\ and\ \citenamefont
  {Danshita}}]{Yamamoto2015}%
  \BibitemOpen
  \bibfield  {author} {\bibinfo {author} {\bibfnamefont {D.}~\bibnamefont
  {Yamamoto}}, \bibinfo {author} {\bibfnamefont {G.}~\bibnamefont
  {Marmorini}},\ and\ \bibinfo {author} {\bibfnamefont {I.}~\bibnamefont
  {Danshita}},\ }\bibfield  {title} {\bibinfo {title} {Microscopic model
  calculations for the magnetization process of layered triangular-lattice
  quantum antiferromagnets},\ }\href
  {https://doi.org/10.1103/PhysRevLett.114.027201} {\bibfield  {journal}
  {\bibinfo  {journal} {Phys. Rev. Lett.}\ }\textbf {\bibinfo {volume} {114}},\
  \bibinfo {pages} {027201} (\bibinfo {year} {2015})}\BibitemShut {NoStop}%
\bibitem [{\citenamefont {Li}\ \emph {et~al.}(2019)\citenamefont {Li},
  \citenamefont {Zelenskiy}, \citenamefont {Quilliam}, \citenamefont {Dun},
  \citenamefont {Zhou}, \citenamefont {Plumer},\ and\ \citenamefont
  {Quirion}}]{Li2019}%
  \BibitemOpen
  \bibfield  {author} {\bibinfo {author} {\bibfnamefont {M.}~\bibnamefont
  {Li}}, \bibinfo {author} {\bibfnamefont {A.}~\bibnamefont {Zelenskiy}},
  \bibinfo {author} {\bibfnamefont {J.~A.}\ \bibnamefont {Quilliam}}, \bibinfo
  {author} {\bibfnamefont {Z.~L.}\ \bibnamefont {Dun}}, \bibinfo {author}
  {\bibfnamefont {H.~D.}\ \bibnamefont {Zhou}}, \bibinfo {author}
  {\bibfnamefont {M.~L.}\ \bibnamefont {Plumer}},\ and\ \bibinfo {author}
  {\bibfnamefont {G.}~\bibnamefont {Quirion}},\ }\bibfield  {title} {\bibinfo
  {title} {Magnetoelastic coupling and the magnetization plateau in
  {${\mathrm{Ba}}_{3}{\mathrm{CoSb}}_{2}{\mathrm{O}}_{9}$}},\ }\href
  {https://doi.org/10.1103/PhysRevB.99.094408} {\bibfield  {journal} {\bibinfo
  {journal} {Phys. Rev. B}\ }\textbf {\bibinfo {volume} {99}},\ \bibinfo
  {pages} {094408} (\bibinfo {year} {2019})}\BibitemShut {NoStop}%
\bibitem [{\citenamefont {Liu}\ \emph {et~al.}(2019)\citenamefont {Liu},
  \citenamefont {Prokhnenko}, \citenamefont {Yamamoto}, \citenamefont
  {Bartkowiak}, \citenamefont {Kurita},\ and\ \citenamefont
  {Tanaka}}]{Liu2019a}%
  \BibitemOpen
  \bibfield  {author} {\bibinfo {author} {\bibfnamefont {X.~Z.}\ \bibnamefont
  {Liu}}, \bibinfo {author} {\bibfnamefont {O.}~\bibnamefont {Prokhnenko}},
  \bibinfo {author} {\bibfnamefont {D.}~\bibnamefont {Yamamoto}}, \bibinfo
  {author} {\bibfnamefont {M.}~\bibnamefont {Bartkowiak}}, \bibinfo {author}
  {\bibfnamefont {N.}~\bibnamefont {Kurita}},\ and\ \bibinfo {author}
  {\bibfnamefont {H.}~\bibnamefont {Tanaka}},\ }\bibfield  {title} {\bibinfo
  {title} {Microscopic evidence of a quantum magnetization process in the
  {$S=\frac{1}{2}$} triangular-lattice {Heisenberg}-like antiferromagnet
  {${\mathrm{Ba}}_{3}{\mathrm{CoSb}}_{2}{\mathrm{O}}_{9}$}},\ }\href
  {https://doi.org/10.1103/PhysRevB.100.094436} {\bibfield  {journal} {\bibinfo
   {journal} {Phys. Rev. B}\ }\textbf {\bibinfo {volume} {100}},\ \bibinfo
  {pages} {094436} (\bibinfo {year} {2019})}\BibitemShut {NoStop}%
\bibitem [{\citenamefont {Fortune}\ \emph {et~al.}(2021)\citenamefont
  {Fortune}, \citenamefont {Huang}, \citenamefont {Hong}, \citenamefont {Ma},
  \citenamefont {Choi}, \citenamefont {Hannahs}, \citenamefont {Zhao},
  \citenamefont {Sun}, \citenamefont {Takano},\ and\ \citenamefont
  {Zhou}}]{Fortune2021}%
  \BibitemOpen
  \bibfield  {author} {\bibinfo {author} {\bibfnamefont {N.~A.}\ \bibnamefont
  {Fortune}}, \bibinfo {author} {\bibfnamefont {Q.}~\bibnamefont {Huang}},
  \bibinfo {author} {\bibfnamefont {T.}~\bibnamefont {Hong}}, \bibinfo {author}
  {\bibfnamefont {J.}~\bibnamefont {Ma}}, \bibinfo {author} {\bibfnamefont
  {E.~S.}\ \bibnamefont {Choi}}, \bibinfo {author} {\bibfnamefont {S.~T.}\
  \bibnamefont {Hannahs}}, \bibinfo {author} {\bibfnamefont {Z.~Y.}\
  \bibnamefont {Zhao}}, \bibinfo {author} {\bibfnamefont {X.~F.}\ \bibnamefont
  {Sun}}, \bibinfo {author} {\bibfnamefont {Y.}~\bibnamefont {Takano}},\ and\
  \bibinfo {author} {\bibfnamefont {H.~D.}\ \bibnamefont {Zhou}},\ }\bibfield
  {title} {\bibinfo {title} {Evolution of magnetic field induced ordering in
  the layered quantum {Heisenberg} triangular-lattice antiferromagnet
  {${\mathrm{Ba}}_{3}{\mathrm{CoSb}}_{2}{\mathrm{O}}_{9}$}},\ }\href
  {https://doi.org/10.1103/PhysRevB.103.184425} {\bibfield  {journal} {\bibinfo
   {journal} {Phys. Rev. B}\ }\textbf {\bibinfo {volume} {103}},\ \bibinfo
  {pages} {184425} (\bibinfo {year} {2021})}\BibitemShut {NoStop}%
\bibitem [{\citenamefont {Ghioldi}\ \emph {et~al.}(2018)\citenamefont
  {Ghioldi}, \citenamefont {Gonzalez}, \citenamefont {Zhang}, \citenamefont
  {Kamiya}, \citenamefont {Manuel}, \citenamefont {Trumper},\ and\
  \citenamefont {Batista}}]{Ghioldi2018}%
  \BibitemOpen
  \bibfield  {author} {\bibinfo {author} {\bibfnamefont {E.~A.}\ \bibnamefont
  {Ghioldi}}, \bibinfo {author} {\bibfnamefont {M.~G.}\ \bibnamefont
  {Gonzalez}}, \bibinfo {author} {\bibfnamefont {S.-S.}\ \bibnamefont {Zhang}},
  \bibinfo {author} {\bibfnamefont {Y.}~\bibnamefont {Kamiya}}, \bibinfo
  {author} {\bibfnamefont {L.~O.}\ \bibnamefont {Manuel}}, \bibinfo {author}
  {\bibfnamefont {A.~E.}\ \bibnamefont {Trumper}},\ and\ \bibinfo {author}
  {\bibfnamefont {C.~D.}\ \bibnamefont {Batista}},\ }\bibfield  {title}
  {\bibinfo {title} {Dynamical structure factor of the triangular
  antiferromagnet: {Schwinger} boson theory beyond mean field},\ }\href
  {https://doi.org/10.1103/PhysRevB.98.184403} {\bibfield  {journal} {\bibinfo
  {journal} {Phys. Rev. B}\ }\textbf {\bibinfo {volume} {98}},\ \bibinfo
  {pages} {184403} (\bibinfo {year} {2018})}\BibitemShut {NoStop}%
\bibitem [{\citenamefont {Zhang}\ and\ \citenamefont {Li}(2020)}]{Zhang2020}%
  \BibitemOpen
  \bibfield  {author} {\bibinfo {author} {\bibfnamefont {C.}~\bibnamefont
  {Zhang}}\ and\ \bibinfo {author} {\bibfnamefont {T.}~\bibnamefont {Li}},\
  }\bibfield  {title} {\bibinfo {title} {Resonating valence bond theory of
  anomalous spin dynamics of spin-$\frac{1}{2}$ triangular lattice {Heisenberg}
  antiferromagnet and its application to
  {${\mathrm{Ba}}_{3}{\mathrm{CoSb}}_{2}{\mathrm{O}}_{9}$}},\ }\href
  {https://doi.org/10.1103/PhysRevB.102.075108} {\bibfield  {journal} {\bibinfo
   {journal} {Phys. Rev. B}\ }\textbf {\bibinfo {volume} {102}},\ \bibinfo
  {pages} {075108} (\bibinfo {year} {2020})}\BibitemShut {NoStop}%
\bibitem [{\citenamefont {Ferrari}\ and\ \citenamefont
  {Becca}(2019)}]{Ferrari2019}%
  \BibitemOpen
  \bibfield  {author} {\bibinfo {author} {\bibfnamefont {F.}~\bibnamefont
  {Ferrari}}\ and\ \bibinfo {author} {\bibfnamefont {F.}~\bibnamefont
  {Becca}},\ }\bibfield  {title} {\bibinfo {title} {Dynamical structure factor
  of the {${J}_{1}\ensuremath{-}{J}_{2}$} {Heisenberg} model on the triangular
  lattice: Magnons, spinons, and gauge fields},\ }\href
  {https://doi.org/10.1103/PhysRevX.9.031026} {\bibfield  {journal} {\bibinfo
  {journal} {Phys. Rev. X}\ }\textbf {\bibinfo {volume} {9}},\ \bibinfo {pages}
  {031026} (\bibinfo {year} {2019})}\BibitemShut {NoStop}%
\bibitem [{\citenamefont {Verresen}\ \emph {et~al.}(2019)\citenamefont
  {Verresen}, \citenamefont {Moessner},\ and\ \citenamefont
  {Pollmann}}]{Verresen2019}%
  \BibitemOpen
  \bibfield  {author} {\bibinfo {author} {\bibfnamefont {R.}~\bibnamefont
  {Verresen}}, \bibinfo {author} {\bibfnamefont {R.}~\bibnamefont {Moessner}},\
  and\ \bibinfo {author} {\bibfnamefont {F.}~\bibnamefont {Pollmann}},\
  }\bibfield  {title} {\bibinfo {title} {Avoided quasiparticle decay from
  strong quantum interactions},\ }\href
  {https://doi.org/10.1038/s41567-019-0535-3} {\bibfield  {journal} {\bibinfo
  {journal} {Nat. Phys.}\ }\textbf {\bibinfo {volume} {15}},\ \bibinfo {pages}
  {750} (\bibinfo {year} {2019})}\BibitemShut {NoStop}%
\bibitem [{\citenamefont {Ivanov}\ \emph {et~al.}(2010)\citenamefont {Ivanov},
  \citenamefont {Nordblad}, \citenamefont {Mathieu}, \citenamefont {Tellgren},\
  and\ \citenamefont {Ritter}}]{Ivanov2010}%
  \BibitemOpen
  \bibfield  {author} {\bibinfo {author} {\bibfnamefont {S.~A.}\ \bibnamefont
  {Ivanov}}, \bibinfo {author} {\bibfnamefont {P.}~\bibnamefont {Nordblad}},
  \bibinfo {author} {\bibfnamefont {R.}~\bibnamefont {Mathieu}}, \bibinfo
  {author} {\bibfnamefont {R.}~\bibnamefont {Tellgren}},\ and\ \bibinfo
  {author} {\bibfnamefont {C.}~\bibnamefont {Ritter}},\ }\bibfield  {title}
  {\bibinfo {title} {Neutron diffraction studies and the magnetism of an
  ordered perovskite: {$\mathrm{Ba}_2\mathrm{CoTeO}_6$}},\ }\href
  {https://doi.org/10.1039/B927498G} {\bibfield  {journal} {\bibinfo  {journal}
  {Dalton Trans.}\ }\textbf {\bibinfo {volume} {39}},\ \bibinfo {pages} {5490}
  (\bibinfo {year} {2010})}\BibitemShut {NoStop}%
\bibitem [{\citenamefont {Chanlert}\ \emph {et~al.}(2016)\citenamefont
  {Chanlert}, \citenamefont {Kurita}, \citenamefont {Tanaka}, \citenamefont
  {Goto}, \citenamefont {Matsuo},\ and\ \citenamefont {Kindo}}]{Chanlert2016}%
  \BibitemOpen
  \bibfield  {author} {\bibinfo {author} {\bibfnamefont {P.}~\bibnamefont
  {Chanlert}}, \bibinfo {author} {\bibfnamefont {N.}~\bibnamefont {Kurita}},
  \bibinfo {author} {\bibfnamefont {H.}~\bibnamefont {Tanaka}}, \bibinfo
  {author} {\bibfnamefont {D.}~\bibnamefont {Goto}}, \bibinfo {author}
  {\bibfnamefont {A.}~\bibnamefont {Matsuo}},\ and\ \bibinfo {author}
  {\bibfnamefont {K.}~\bibnamefont {Kindo}},\ }\bibfield  {title} {\bibinfo
  {title} {Field-driven successive phase transitions in the
  quasi-two-dimensional frustrated antiferromagnet
  {${\mathrm{Ba}}_{2}{\mathrm{CoTeO}}_{6}$} and highly degenerate classical
  ground states},\ }\href {https://doi.org/10.1103/PhysRevB.93.094420}
  {\bibfield  {journal} {\bibinfo  {journal} {Phys. Rev. B}\ }\textbf {\bibinfo
  {volume} {93}},\ \bibinfo {pages} {094420} (\bibinfo {year}
  {2016})}\BibitemShut {NoStop}%
\bibitem [{\citenamefont {Chanlert}\ \emph {et~al.}(2017)\citenamefont
  {Chanlert}, \citenamefont {Kurita}, \citenamefont {Tanaka}, \citenamefont
  {Kimata},\ and\ \citenamefont {Nojiri}}]{Chanlert2017}%
  \BibitemOpen
  \bibfield  {author} {\bibinfo {author} {\bibfnamefont {P.}~\bibnamefont
  {Chanlert}}, \bibinfo {author} {\bibfnamefont {N.}~\bibnamefont {Kurita}},
  \bibinfo {author} {\bibfnamefont {H.}~\bibnamefont {Tanaka}}, \bibinfo
  {author} {\bibfnamefont {M.}~\bibnamefont {Kimata}},\ and\ \bibinfo {author}
  {\bibfnamefont {H.}~\bibnamefont {Nojiri}},\ }\bibfield  {title} {\bibinfo
  {title} {Collective and local excitations in
  {${\mathrm{Ba}}_{2}{\mathrm{CoTeO}}_{6}$}: A composite system of a spin-1/2
  triangular-lattice {Heisenberg} antiferromagnet and a honeycomb-lattice
  {${J}_{1}\ensuremath{-}{J}_{2}$} {Ising} antiferromagnet},\ }\href
  {https://doi.org/10.1103/PhysRevB.96.064419} {\bibfield  {journal} {\bibinfo
  {journal} {Phys. Rev. B}\ }\textbf {\bibinfo {volume} {96}},\ \bibinfo
  {pages} {064419} (\bibinfo {year} {2017})}\BibitemShut {NoStop}%
\bibitem [{\citenamefont {Nakajima}\ \emph {et~al.}(2011)\citenamefont
  {Nakajima}, \citenamefont {Ohira-Kawamura}, \citenamefont {Kikuchi},
  \citenamefont {Nakamura}, \citenamefont {Kajimoto}, \citenamefont {Inamura},
  \citenamefont {Takahashi}, \citenamefont {Aizawa}, \citenamefont {Suzuya},
  \citenamefont {Shibata}, \citenamefont {Nakatani}, \citenamefont {Soyama},
  \citenamefont {Maruyama}, \citenamefont {Tanaka}, \citenamefont {Kambara},
  \citenamefont {Iwahashi}, \citenamefont {Itoh}, \citenamefont {Osakabe},
  \citenamefont {Wakimoto}, \citenamefont {Kakurai}, \citenamefont {Maekawa},
  \citenamefont {Harada}, \citenamefont {Oikawa}, \citenamefont {E.~Lechner},
  \citenamefont {Mezei},\ and\ \citenamefont {Arai}}]{Nakajima2011}%
  \BibitemOpen
  \bibfield  {author} {\bibinfo {author} {\bibfnamefont {K.}~\bibnamefont
  {Nakajima}}, \bibinfo {author} {\bibfnamefont {S.}~\bibnamefont
  {Ohira-Kawamura}}, \bibinfo {author} {\bibfnamefont {T.}~\bibnamefont
  {Kikuchi}}, \bibinfo {author} {\bibfnamefont {M.}~\bibnamefont {Nakamura}},
  \bibinfo {author} {\bibfnamefont {R.}~\bibnamefont {Kajimoto}}, \bibinfo
  {author} {\bibfnamefont {Y.}~\bibnamefont {Inamura}}, \bibinfo {author}
  {\bibfnamefont {N.}~\bibnamefont {Takahashi}}, \bibinfo {author}
  {\bibfnamefont {K.}~\bibnamefont {Aizawa}}, \bibinfo {author} {\bibfnamefont
  {K.}~\bibnamefont {Suzuya}}, \bibinfo {author} {\bibfnamefont
  {K.}~\bibnamefont {Shibata}}, \bibinfo {author} {\bibfnamefont
  {T.}~\bibnamefont {Nakatani}}, \bibinfo {author} {\bibfnamefont
  {K.}~\bibnamefont {Soyama}}, \bibinfo {author} {\bibfnamefont
  {R.}~\bibnamefont {Maruyama}}, \bibinfo {author} {\bibfnamefont
  {H.}~\bibnamefont {Tanaka}}, \bibinfo {author} {\bibfnamefont
  {W.}~\bibnamefont {Kambara}}, \bibinfo {author} {\bibfnamefont
  {T.}~\bibnamefont {Iwahashi}}, \bibinfo {author} {\bibfnamefont
  {Y.}~\bibnamefont {Itoh}}, \bibinfo {author} {\bibfnamefont {T.}~\bibnamefont
  {Osakabe}}, \bibinfo {author} {\bibfnamefont {S.}~\bibnamefont {Wakimoto}},
  \bibinfo {author} {\bibfnamefont {K.}~\bibnamefont {Kakurai}}, \bibinfo
  {author} {\bibfnamefont {F.}~\bibnamefont {Maekawa}}, \bibinfo {author}
  {\bibfnamefont {M.}~\bibnamefont {Harada}}, \bibinfo {author} {\bibfnamefont
  {K.}~\bibnamefont {Oikawa}}, \bibinfo {author} {\bibfnamefont
  {R.}~\bibnamefont {E.~Lechner}}, \bibinfo {author} {\bibfnamefont
  {F.}~\bibnamefont {Mezei}},\ and\ \bibinfo {author} {\bibfnamefont
  {M.}~\bibnamefont {Arai}},\ }\bibfield  {title} {\bibinfo {title}
  {{AMATERAS}: A cold-neutron disk chopper spectrometer},\ }\href
  {https://doi.org/10.1143/JPSJS.80SB.SB028} {\bibfield  {journal} {\bibinfo
  {journal} {J. Phys. Soc. Jpn.}\ }\textbf {\bibinfo {volume} {80}},\ \bibinfo
  {pages} {SB028} (\bibinfo {year} {2011})}\BibitemShut {NoStop}%
\bibitem [{\citenamefont {Inamura}\ \emph {et~al.}(2013)\citenamefont
  {Inamura}, \citenamefont {Nakatani}, \citenamefont {Suzuki},\ and\
  \citenamefont {Otomo}}]{Inamura2013}%
  \BibitemOpen
  \bibfield  {author} {\bibinfo {author} {\bibfnamefont {Y.}~\bibnamefont
  {Inamura}}, \bibinfo {author} {\bibfnamefont {T.}~\bibnamefont {Nakatani}},
  \bibinfo {author} {\bibfnamefont {J.}~\bibnamefont {Suzuki}},\ and\ \bibinfo
  {author} {\bibfnamefont {T.}~\bibnamefont {Otomo}},\ }\bibfield  {title}
  {\bibinfo {title} {Development status of software ``{Utsusemi}'' for chopper
  spectrometers at {MLF}, {J-PARC}},\ }\href
  {https://doi.org/10.7566/JPSJS.82SA.SA031} {\bibfield  {journal} {\bibinfo
  {journal} {J. Phys. Soc. Jpn.}\ }\textbf {\bibinfo {volume} {82}},\ \bibinfo
  {pages} {SA031} (\bibinfo {year} {2013})}\BibitemShut {NoStop}%
\end{thebibliography}

\providecommand{\noopsort}[1]{}\providecommand{\singleletter}[1]{#1}%

\end{document}